\documentclass[prx,twocolumn,amsmath,amssymb,showpacs,longbibliography]{revtex4-1}
\usepackage{graphicx}
\usepackage{color}
\usepackage{float}
\usepackage{hhline}
\usepackage{bm}
\usepackage{pifont}
\usepackage{graphicx}
\usepackage{epstopdf}
\usepackage{amsmath}
\usepackage{amssymb}
\usepackage{booktabs}
\usepackage{color}

\begin{document}

\title{Roles of energy dissipation in a liquid-solid transition \\ of out-of-equilibrium systems}

\author{Yuta Komatsu}
\author{Hajime Tanaka}
\email{tanaka@iis.u-tokyo.ac.jp}
\affiliation{ {Institute of Industrial Science, University of Tokyo, 4-6-1 Komaba, Meguro-ku, Tokyo 153-8505, Japan} }

\date{Received February 18, 2015}

\begin{abstract}
Self-organization of active matter as well as driven granular matter in non-equilibrium dynamical states has attracted considerable attention not only from the fundamental and application viewpoints but also as a model to understand the occurrence of such phenomena in nature. These systems share common features originating from their intrinsically out-of-equilibrium nature. It remains elusive how energy dissipation affects the state selection in such non-equilibrium states. As a simple model system, we consider a non-equilibrium stationary state maintained by continuous energy input, relevant to industrial processing of granular materials by vibration and/or flow. More specifically, we experimentally study roles of dissipation in self-organization of a driven granular particle monolayer. We find that the introduction of strong inelasticity entirely changes the nature of the liquid-solid transition from two-step (nearly) continuous transitions (liquid-hexatic-solid)  to a strongly discontinuous first-order-like one (liquid-solid), where the two phases with different effective temperatures can coexist, unlike thermal systems, under a balance between energy input and dissipation. Our finding indicates a pivotal role of energy dissipation and suggests a novel principle in the self-organization of  systems far from equilibrium. A similar principle may apply to active matter, which is another important class of out-of-equilibrium systems. On noting that interaction forces in active matter, and particularly in living systems, are often non-conservative and dissipative, our finding may also shed new light on the state selection in these systems.  
\end{abstract}

\maketitle

\section{Introduction}
Despite the fact that self-organization of a system in an out-of-equilibrium state plays a crucial role in dynamical structural formation in nature, 
physical principles behind such phenomena have remained elusive. 
Active matter \cite{marchetti2013hydrodynamics} and driven granular matter \cite{jaeger1996granular,kadanoff1999,aranson2006patterns} are two important classes of 
out-of-equilibrium systems. 
They share an intrinsic out-of-equilibrium nature, and the only basic difference is that the energy is injected locally for the active systems whereas 
globally for the granular systems \cite{marchetti2013hydrodynamics}. This global nature of energy input makes granular matter physically simpler than active matter. 
Thus, granular matter is not only important for its own sake, but also regarded as a model for understanding the physics of active matter. 

Granular matter is an important class of materials, distinct from thermal systems 
since the thermal energy is negligible for its description. 
Granular matter is ubiquitous in nature and its dynamical self-organization always takes place in a strongly non-equilibrium 
situation as in active matter, since energy input is essential for its occurrence \cite{jaeger1996granular,kadanoff1999,aranson2006patterns}. 
Its statistical yet athermal nature makes the physical description extremely difficult. 
From an experimental point of view, the control of self-organization of granular matter is also a difficult task. 
However, a notable exception is a dynamic steady state, maintained by the balance between energy input and dissipation, 
which allows us to perform well-controlled experiments.  
The most idealized system may be a quasi-two-dimensional (2D) driven granular particle monolayer, where spherical particles are 
confined between two parallel plates whose gap is narrow enough to avoid particle overlap along the vertical direction 
and energy is injected by vertically vibrating plates. This system allows us to access 
all phase-space information at the particle level. 
So the phase behavior of such a monolayer particle system has played a crucial role in our understanding of the 
fundamental nature of self-organization in a system far from equilibrium. 

This vibrated monolayer particle system has also attracted considerable attention for its connections with fundamental problems in the field of 
condensed matter and statistical physics \cite{NelsonB}. 
The liquid-solid transition in a 2D disk system, the thermodynamic counterpart of a vibrated monolayer, has been a hot topic  
since the discovery of the liquid-solid transition for hard disks by Alder and Wainwright \cite{alder1962phase}. 
Two-dimensional particle systems cannot crystallize
at finite temperature due to significant fluctuation effects associated with the low dimensionality, yet the above work shows 
that they may still form solids.
There is a long-standing debate \cite{2Dmelting} on the nature of this transition for a system of the simplest interparticle interaction, hard disks. 
One scenario is that ordering takes place via two steps, liquid-to-hexatic and hexatic-to-solid transitions, now widely known as 
the Kosterlitz-Thouless-Halperin-Nelson-Young (KTHNY) scenario \cite{NelsonB,kosterlitz1973ordering,Halperin1978,Young1979}. Here each transition is continuous. 
The other is that ordering takes place in one step via a first-order liquid-solid transition \cite{chui1983}. 
There have been hot debates on which is the relevant scenario. Very recently, it was shown \cite{bernard2011two} that the transition 
actually takes place by a scenario different from both of them: It occurs with two steps as the KTHNY scenario suggests, 
but the first transition is not continuous but weakly discontinuous. 
However, the first-order nature of the liquid-hexatic transition is very weak, and the transition roughly obeys the 
KTHNY scenario. This basic behavior is common to other systems including particles interacting with soft repulsive potentials \cite{keim2010melting,wang2011two,mazars2013melting,kapfer2014soft}
and those with attractive potentials such as the Lennard-Jones potential \cite{2DLJ}, although 
it has recently be shown that the nature of the transitions depends on the softness of the potential in a delicate manner \cite{kapfer2014soft}. 
Monolayer granular matter has provided a model experimental system to study this fundamental problem. 

Some time ago, careful experiments were made on the athermal counterpart of the above system. 
It was shown by Shattuck and his coworkers that a driven monolayer particle system continuously transforms from a liquid to an intermediate hexatic, and then to a solid phase, 
with an increase in the particle area fraction $\phi$ under a constant $\Gamma$ \cite{shattuck2006crystallization} (see Sec. II for the precise definitions of $\Gamma$ and $\phi$).  
A similar meting transition behavior was also observed by Olasfen and Urbach when increasing the dimensionless acceleration $\Gamma$ at a fixed 
particle area fraction $\phi$ for a granular quasi-monolayer \cite{olafsen2005two}. However, it was shown that the thickness of the cell $h$, which is 1.6 times of the particle diameter $d$, 
plays a crucial role in the transition: height fluctuations of particles may be a source of disorder. The increase of their amplitude with an increase in the vibration amplitude, or $\Gamma$, 
increases the number density of defects, eventually leading to the melting of the solid phase. Thus, the mechanism may be essentially different from the former example, 
which does not involve any significant hight fluctuations due to a strong 2D confinement.    
 
The former liquid-solid transition behavior as a function of $\phi$ \cite{shattuck2006crystallization} obeys the KTHNY scenario \cite{NelsonB,kosterlitz1973ordering,Halperin1978,Young1979}, although 
the liquid-to-hexatic transition may be weakly first-order \cite{bernard2011two,kapfer2014soft}. 
This study suggests that a quasi-2D driven granular system behaves very similarly to its thermal counterpart.     
A similar conclusion was also derived for glass-transition-like phenomena of driven binary mixtures \cite{durian2007measurement} and polydisperse systems \cite{Watanabe,watanabe2011}. 
The energy injected by mechanical vibration is converted to the kinetic energy of a system and the effective (granular) temperature $T^\ast$ is 
defined by this kinetic energy. If this is high enough to overcome the gravity and the energy loss originating from the friction and inelastic collisions 
with the container \cite{olafsen1998clustering}, we may approximately regard the system as a thermal equilibrium system as long as interparticle collisions are almost elastic. 
However, the exact mechanism responsible for this apparent thermal behavior of an athermal system has remained elusive. 
We note that these experiments were performed by using rather elastic balls such as steel balls. 
Then, the natural question to ask next is how the nature of the liquid-solid transition is affected by the inelasticity of collisions, or internal dissipation. 

There were pioneering works on liquid-solid transitions of quasi-2D granular systems 
\cite{urbach2004,urbach2005dynamics,urbach2008effect,urbach2009effects,luu2013capillarylike,clerc2008liquid}. 
These studies showed interesting monolayer liquid-bilayer solid coexistence in a nonequilibrium steady state, in which the two phases have different granular temperatures. 
These are intriguing examples of phase ordering, more generally self-organization,  
in a dynamic steady state of a non-equilibrium open system, maintained by the balance between energy input and dissipation.

Here we study the effect of energy dissipation on liquid-solid transition of a quasi-2D driven granular matter by comparing the behaviors of steel and rubber ball systems.  
We note that steel and rubber balls have differences in the restitution coefficient, the friction coefficient, and the elastic properties. 
Among these, the difference in elastic properties may be less important compared to the others because forces acting upon interparticle collisions are too weak to cause non-linear deformation of the balls  
and it is known that the 2D melting behavior is not affected by the softness of the interaction potential \cite{kapfer2014soft}.   
A situation we consider is a {\it single} layer of monodisperse spheres, which is vibrated between two horizontal plates \cite{pieranski1978hard}, 
unlike the above-mentioned  previous works where bilayer formation is allowed  \cite{urbach2004,urbach2005dynamics,urbach2008effect,urbach2009effects,luu2013capillarylike,clerc2008liquid}. 
The control parameters in our experiments are the energy input characterized by a dimensionless acceleration 
$\Gamma$ and the area fraction of particles $\phi$. 
Here we demonstrate that the dissipation due to the inelasticity of collisions and the friction can, if they are strong enough, completely break the similarity between 
the athermal and the corresponding thermal system 
and fundamentally change the nature of the liquid-solid transition {\it in a monolayer} from the KTHNY-like continuous transitional behavior \cite{shattuck2006crystallization} 
to a strongly discontinuous transition. 
We discuss a physical mechanism responsible for this unconventional self-organization under energy dissipation. 
Our study reveals a novel mechanism leading to the coexistence of two phases with different granular temperatures, which 
is essentially different from the mechanism previously found \cite{urbach2004,urbach2005dynamics,urbach2008effect,urbach2009effects,luu2013capillarylike,clerc2008liquid}.   
We infer that a similar mechanism may be relevant to self-organization of active matter including living systems. Dissipative interactions, such as 
inelastic, hydrodynamic, frictional  interactions, may play a crucial role in the state selection.   

\section{Experimental}

\subsection{Experimental systems}
Our experimental apparatus is analogous to those used in 
\cite{olafsen2005two,shattuck2006crystallization} and the same as 
that used in \cite{Watanabe,watanabe2011}. 
A monolayer of non-cohesive particles was vibrated sinusoidally in the vertical direction by 
an electromagnetic shaker (Labworks ET-139) with frequency $f=50\,\mathrm{Hz}$. 
We changed the dimensionless acceleration 
$\Gamma=A(2\pi f)^2/g$, where $A$ is the amplitude of 
vibration and $g$ is the gravitational acceleration, by controlling $A$. 
The cell has a circular shape and its inner diameter is $L=102.5\,\mathrm{mm}$. 
The circular annulus made of duralumin is used as the hard side wall. 
When we make an experiment with a soft side wall (see Sec. II.D), the inner side wall of the annulus is covered by a silicone rubber sheet of 0.2 mm thickness. 
In our experiments, we use two types of spherical particles: steel and fluorine-rubber balls. 
The particle diameter is $d=3.0\,\mathrm{mm}$ for both stainless steel and fluorine-rubber ball. 
The fluorine-rubber ball has four important characteristics: (i) a non-cohesive character, (ii) a large stiffness ($\sim 1$ GPa), 
which allows us to ignore nonlinear deformation upon collision, 
(iii) a low dynamic friction coefficient ($<0.4$), and (iv) a low restitution coefficient. 
The friction coefficient of the fluorine rubber may be comparable to that of the steel. However, the coefficient of 
rolling friction may be smaller for the steel ball than for the rubber ball since the former has a smoother surface than the latter.  
 
The restitution coefficient $\alpha$ is difficult to estimate accurately and it is known to depend on the particle velocity. 
Here we measured the height ratio before and after a collision with a steel wall by dropping a ball vertically. 
The restitution coefficient $\alpha$ against a steel wall was estimated as $\sim 0.8$ and $\sim 0.3$ with the variance of 0.1 for the steel and rubber ball respectively. 
We also estimated the $\alpha$ for collision between two rubber balls as $\sim 0.1$. 
The value of $\alpha$ for steel balls coincides well with the literature value \cite{windows2014inelasticity}.   
By confining these particles between two plates separated by an annulus with 
a thickness $h$ of 4.0\,mm (i.e., $h/d \cong 1.33$),
we allowed only quasi-2D particle motion, i.e, no bilayer formation. 
We note that in previous similar works on the liquid-solid transition of driven granular matter \cite{urbach2008effect,luu2013capillarylike} 
$h/d=1.74 \sim 1.95$, which allows particles to form bilayers (see Appendix A-1). 
This difference in $h/d$ leads to a crucial difference in the nature of a liquid-solid transition 
between our system and theirs, as will be discussed later. 

The top plate is a transparent glass plate so that we can observe 
the particles through it. 
The bottom duralumin plate is covered with a sandpaper by which the 
vertically oscillating particles are scattered \cite{shattuck2006crystallization}. 
The use of a wall with surface roughness is expected to randomize the horizontal motion and realize Brownian-like motion of particles. 
It was shown \cite{olafsen1999velocity} that a large enough $\Gamma$ leads to the velocity distribution function of a nearly Gaussian shape 
for a driven monolayer system. 
This is because the randomization of the energy injection due to particle collisions and the wall roughness leads to realization of 
a self-generated effective white bath in a long timescale.

For steel balls, we obtained the particle coordinates by tracking the bright spots at the top of particles 
reflecting illuminating light.  
For rubber balls, on the other hand, we obtain the particle coordinates simply by 
binarisation of particle images. 
We used a high speed camera (VCC-H500, DigiMo Co. Ltd.). The typical frame rate used was 100 frame/s  
and the image resolution was 640$\times$480. Occasionally, we used the rate of 500 frame/s with a resolution of 640$\times$90. A pixel size corresponds to 0.25 mm. 
The position of each particle was tracked by a particle tracking software, which fits a Gaussian function to an image of each particle. 
The detection error is less than 0.05 mm for a steel ball and 0.1 mm for a rubber ball.    
We measured a structure after attaining a steady state (typically after 10 min from the initiation of vibrational driving). 

In our system, the transition between a solid-like and liquid-like state occurs as a function of the area fraction  
$\phi=(N d^2)/L^2$, where $d$ is the diameter of particle, $L=102.5\,\mathrm{mm}$ is the inner diameter of the annulus, $N$ is the total number of particles 
in the system. We note that this definition of $\phi$ is often used in describing the phase transition of  (quasi-)2D hard particle systems (see, e.g., \cite{shattuck2006crystallization}).  
We note that the random close packing and the closest packing in 2D occur at about 0.83 and 0.906 respectively.

\subsection{Finite-size effects}
 
Because of the rather small size of our system, our measurements may suffer from finite-size effects. 
The effects may not be so serious for a discontinuous liquid-solid transition in a rubber ball system, since there is no diverging lengthscale. 
On the other hand, the results of a steel ball system may suffer from finite-size effects. However, the maximum correlation length 
we could attain was 9 times the particle diameter $d$, which is still much shorter than  the cell diameter ($\sim 34d$). 
So the finite size effects may not be so severe for our results, although there might be slight shifts in the transition area fractions. 
This conclusion is supported by the rather good agreement of the measured $\phi$-dependences of the correlation length of hexatic order $\xi_6$ and 
translational order $\xi$ with the theoretical predictions for an infinite system (see below).  
However, the correlation lengths, $\xi_6$ and $\xi$, may be bound by the cell size very near the hexatic ordering point $\phi_h$ and the solidification point $\phi_s$, respectively, 
and the weak discontinuous nature of a liquid-hexatic transition may be smeared out by the finite-size effects \cite{bernard2011two}. 
In principle, we can use a cell with a larger diameter, but what is most important in our measurements is 
the homogeneity of the vertical vibration amplitude.  So we used only the rather small cell. 
We are planning to make a large cell with high mechanical rigidity, but we leave this for future investigation.

\subsection{Characterization of structures}
 
The 2D radial distribution function $g(r)$ was calculated as 
\begin{eqnarray}
g(r)=\frac{1}{2\pi r\Delta r \rho (N-1)}\sum_{j \neq k} 
\delta(r-|\vec{r}_{jk}|), \nonumber 
\end{eqnarray}
which is the ratio of the ensemble average of the number density of particles 
existing in the region $r \sim r+\Delta r$ to the average number density 
$\rho=N/L^2$.  
Here $N$ is the number of particles in the simulation box, 
whose side length is $L$, and $\Delta r$ is the increment of $r$. 
The spatial correlation of translational order is characterized by the translational correlation length $\xi$ for $\phi<\phi_s$ as follows:
\begin{eqnarray}
En^+(g(r)-1) \cong \exp(-r/\xi).  \nonumber
\end{eqnarray}
where $En^+(f(x))$ expresses an operation to extract the envelop function of the positive part of a function $f(x)$. 

The hexatic order parameter is
measured by $\psi_6^j=(1/n_j)\sum_k \exp \left( i6\theta_{jk} \right)$ 
for each particle $j$,
where the sum runs over the $n_j$ nearest-neighbors of particle $j$,
and $\theta_{jk}$ is the angle between the bond $\vec{r}_k-\vec{r}_j$ 
and a fixed arbitrary axis. 
Here the nearest-neighbor particles are detected by a criterion $r<1.4d$. 
We confirmed that this criterion provides the same nearest-neighbor identification as a method based 
on Voronoi tessellation.  

Then, the spatial correlation of $\psi_6^{j}$ is calculated 
as \cite{NelsonB}
\begin{eqnarray}
g_6^{2D}(r)=\frac{L^2}{2\pi r\Delta rN(N-1)}\sum_{j \neq k}\delta(r-|\vec{r}_{jk}|)
\psi_6^{j}\psi_6^{k\ast}. \nonumber
\end{eqnarray}
The spatial correlation of the bond-orientational order 
can then be characterized by $g_6^{2D}(r)/g(r)$.  Here the division by $g(r)$ is to remove the effect of translational ordering. 

To characterize the fluctuations of $\psi_6$, 
we estimate the spatial correlation length of $\psi_6$, $\xi_6$, and the susceptibility, $\chi_6=\langle (\psi_6-\langle \psi_6 \rangle)^2 \rangle$. 
We obtain the spatial correlation length $\xi_6$ for $\phi<\phi_h$ from the following relation: 
\begin{eqnarray}
En^+(g_6^{2D}(r)/g(r)) \cong \exp(-r/\xi_6).  \nonumber
\end{eqnarray}

\section{Results}

\subsection{Ordering in a steel ball system upon densification.}
Now we show experimental data which provide information on the nature of the phase transitions. 
For driven steel balls, it was previously shown that the system transforms from the liquid to solid via the intermediate hexatic phase \cite{shattuck2006crystallization}. 
For comparing the behavior of steel and rubber ball systems on the same ground we performed experiments for the two systems  
under the same experimental conditions. 

\begin{figure}[h!]
  \begin{center}
   \includegraphics[width =8.5cm]{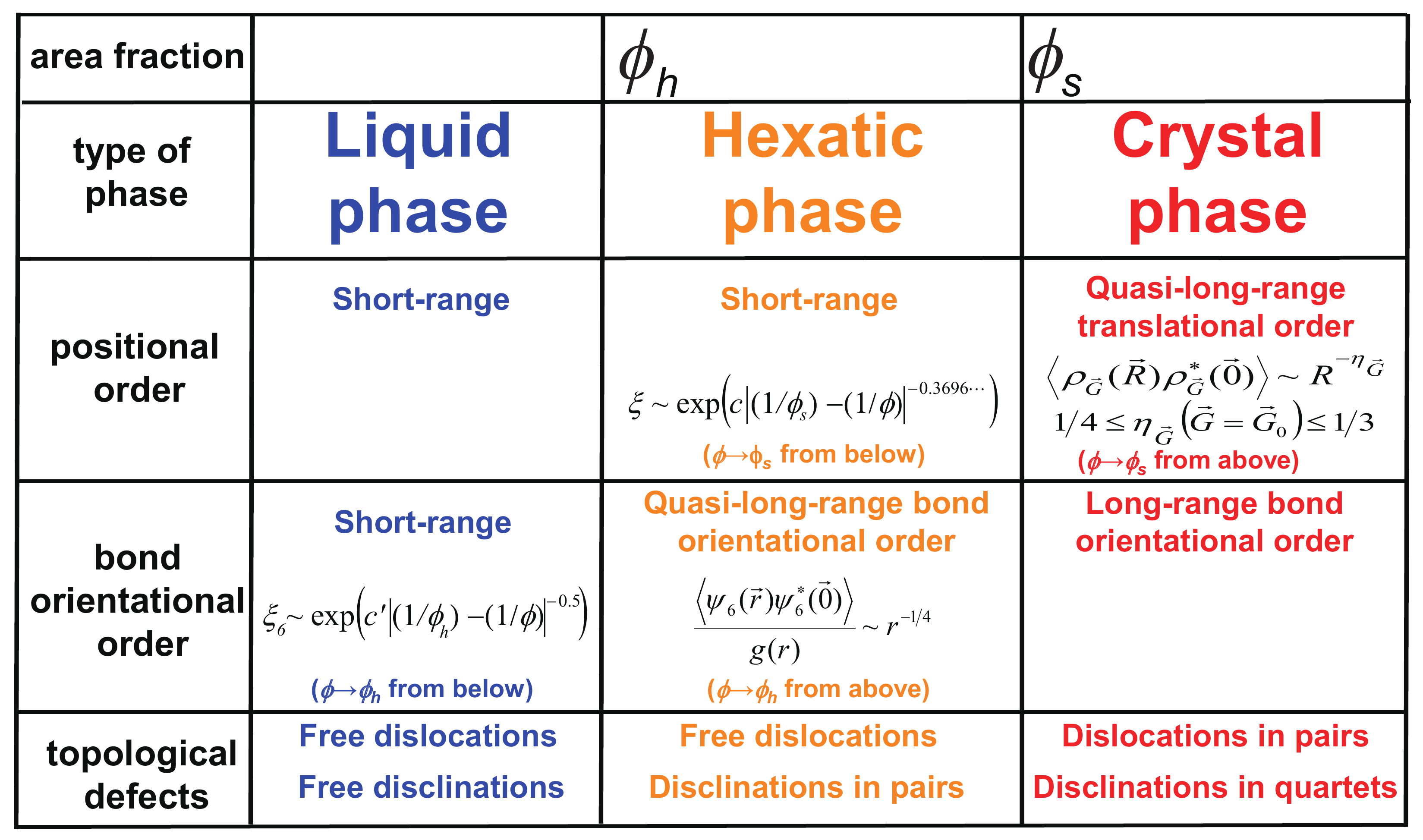}
  \end{center}
  \caption{Phase behavior of 2D hard disks predicted by the KTHNY theory. 
  In this table, we summarize how the spatial correlations of the positional and bond orientational order parameter 
should increase with an increase in $\phi$. The other characteristics of each phase are also shown. } 
\label{fig:KTHNY}
\end{figure}

Before showing results, we briefly review how the spatial correlations of the positional and bond orientational order parameter 
are predicted to grow with an increase in the area fraction $\phi$ in the framework of the KTHNY theory \cite{NelsonB}.  
The predicted behaviors are summarized in Fig. \ref{fig:KTHNY}. 
The hexatic ordering point $\phi_{h}$ and the crystallization point $\phi_{s}$ are characterized by the power-law decays 
of the bond orientational and positional order parameter, respectively. The former should obey $r^{-1/4}$ at $\phi_{h}$, whereas  
the latter $r^{-1/3}$ at $\phi_s$.

\begin{figure}[h!]
\begin{center}
   \includegraphics[width =8.cm]{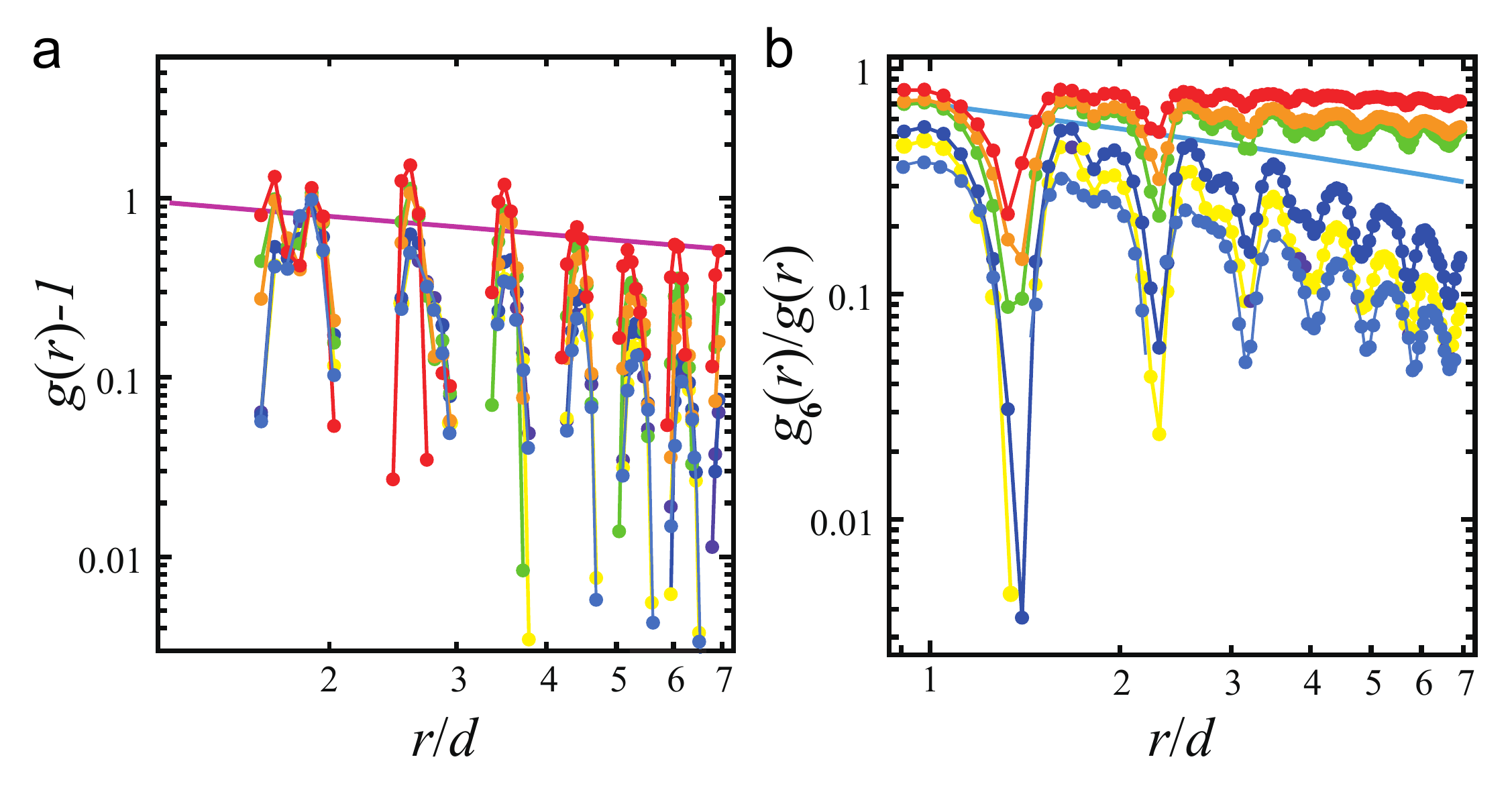}
\end{center}
\caption{$\phi$-dependence of $g(r)$ and $g_6(r)/g(r)$ for a steel ball system for $\Gamma=3.3$.  
To visually confirm a decay slower than the asymptotic decay for ordered phases, we plot the data in the log-log plots, 
where only the points having positive values are displayed.   
(a) $\phi$-dependence of $g(r)$. The signals are $\phi=$0.70 (dark blue), 0.71 (yellow), 0.72 (green), 0.73 (orange), and 0.74 (red)
from the bottom to the top. The straight line has a slope of -1/3. 
(b) $\phi$-dependence of $g_6(r)/g(r)$. 
The signals are $\phi=$0.65 (right blue), 0.70 (dark blue), 0.71 (yellow), 0.72 (green), 0.73 (orange), and 0.74 (red) 
from the bottom to the top. The straight line has a slope of -1/4.  
From these results, we identify $\phi=$0.65, 0.70 and 0.71 as a liquid state, $\phi=$0.72 and 0.73 as 
a hexatic state, and $\phi=$0.74 as a solid state. 
  }
  \label{fig:one}
\end{figure}

We first describe the results of a steel ball system.  
We show the $\phi$-dependence of $g(r)$ and $g_6(r)/g(r)$ for a steel ball system in Fig. \ref{fig:one}. 
The spatial decays of both quantities change from an exponential to a power-law decay, we identify 
the ordering points by whether the decay is slower than the predicted power law decay or not. 
Because of the limitation of our system size, the firm confirmation of the exponent of 
asymptotic power-law decay of these correlation functions are difficult. 
But the results are at least consistent with the prediction of the KTHNY scenario (see also below). 
As shown in Fig. \ref{fig:KTHNY}, $En^+(g_6^{2D}(r)/g(r))$ should decay slower than $r^{-1/4}$ for $\phi \geq \phi_h$, whereas 
$En^+(g(r))$ should decay slower than $r^{-1/3}$ for $\phi \geq \phi_s$. In Figs. \ref{fig:one}(a) and (b), these lines with a slope of $-1/4$ and $-1/3$ 
are drawn as guides to estimate $\phi_{h}$ and $\phi_{s}$ respectively.

In this way we determine $\phi_{h} \sim 0.72$ and $\phi_{s} \sim 0.735$. 
The determinations of $\phi_{h}$ and $\phi_{s}$ are further supported by 
the divergence of the correlation length and the susceptibility of the hexatic order parameter for the former 
and by the divergence of the translational order correlation length for the latter.   
The slight differences of the transition area fractions from those of a hard disk system \cite{bernard2011two} 
may be due to the quasi-2D nature of a system and finite size effects.

\begin{figure}[t!]
  \begin{center}
   \includegraphics[width =7.5cm]{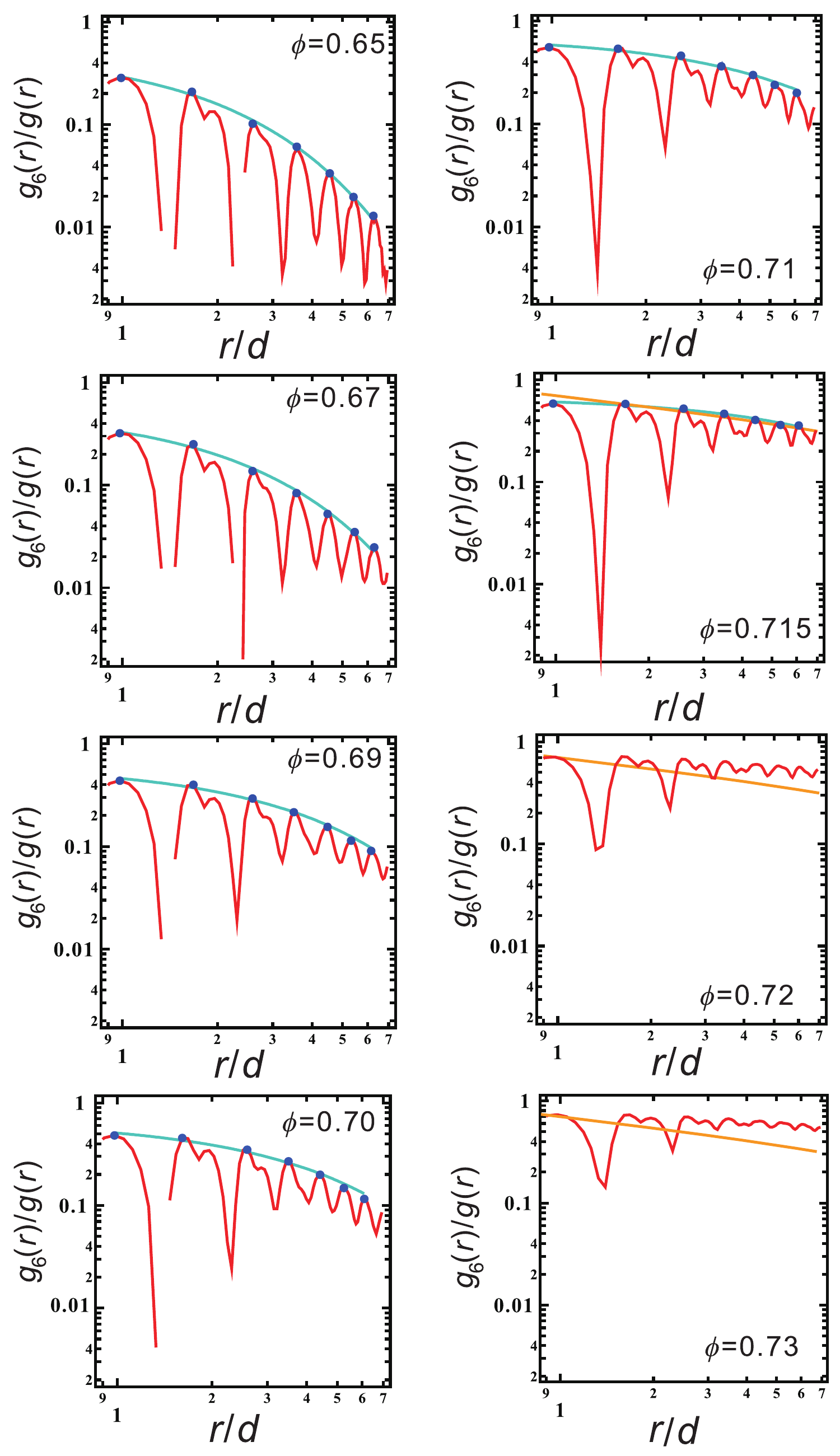}
  \end{center}
  \caption{Examples of the analysis of the spatial decay of $g_6(r)/g(r)$ for a steel ball system for $\Gamma=3.3$.  
  The blue points are used for the fittings. 
The blue curves are exponential fittings, whereas the orange lines have a slope of -1/4, which is expected at $\phi_h$.  From the above, we can judge that 
the liquid-to-hexatic transition takes place between $\phi=0.715$ and 0.72. 
The accuracy of this determination is limited by the discreteness of the area fractions we employed. 
}
  \label{fig:supple_xi6}
  \vspace{3mm}
\end{figure}

\begin{figure}[h!]
\begin{center}
\includegraphics[width=8.cm]{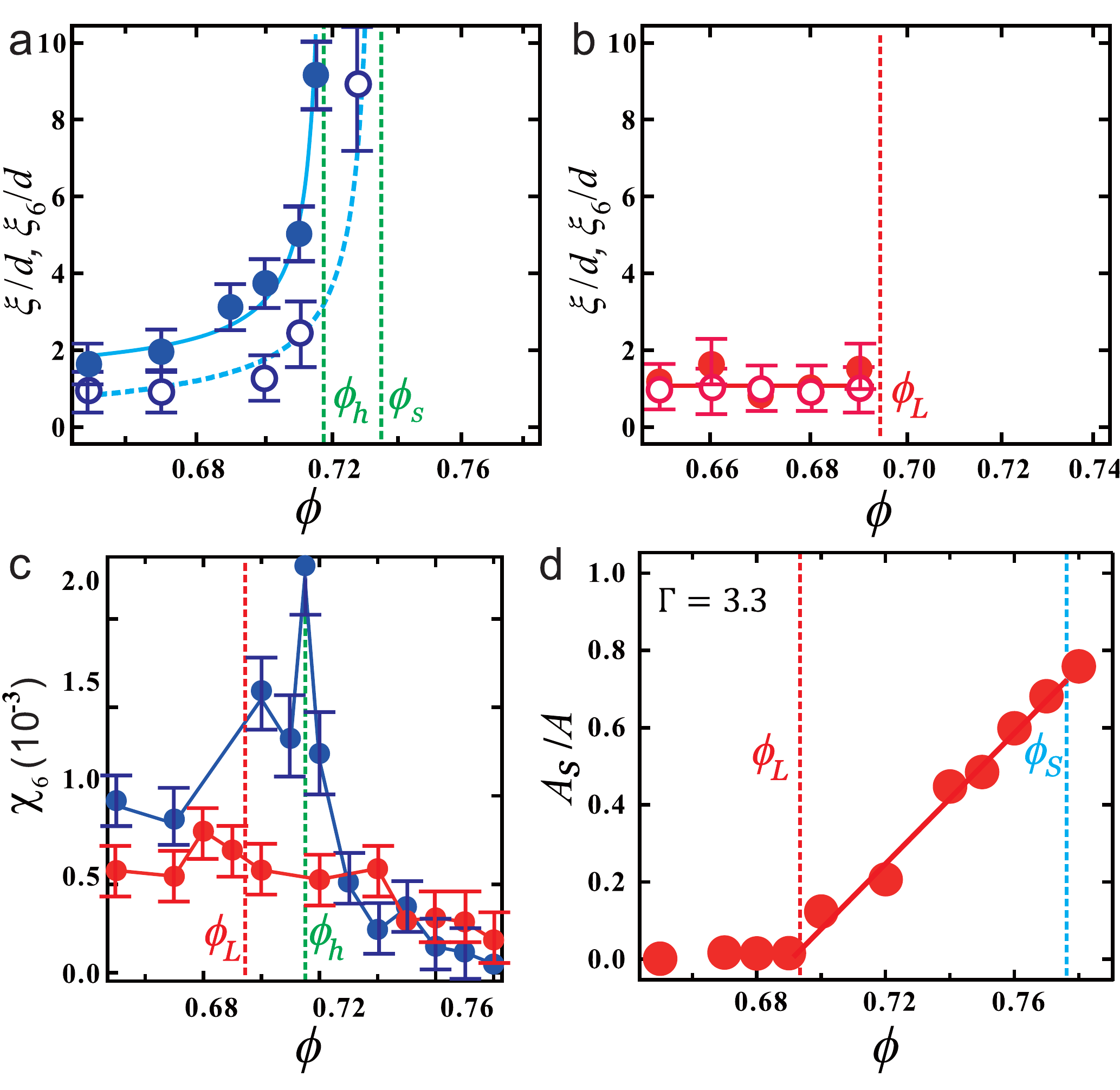}
\end{center}
\caption{Continuous and discontinuous nature of the phase transition respectively in steel and rubber ball systems.  
Blue curves and symbols are for a steel ball system, whereas red ones are for a rubber ball system. 
(a) The divergence of the correlation length of the hexatic order, $\xi_6$, (filled circle) towards 
the hexatic ordering point $\phi_h$ and that of the translational order, $\xi$,  (open circle) towards the solidification point $\phi_s$ 
for a steel ball system. The solid and dashed curves are the prediction of the KTHNY theory 
(see Fig. \ref{fig:KTHNY}). 
(b) The $\phi$-dependence of $\xi_6$ (filled circle) and $\xi$ (open circle) for a homogeneous liquid state of a rubber ball system. 
(c) The susceptibility of the hexatic order parameter $\chi_6$ for steel (blue filled circle) and rubber ball systems (red filled circle). 
The steep increase near $\phi_h$ is observed for a steel ball system, reflecting the nearly continuous nature 
of the transition, whereas there is no such behavior for a rubber ball system, reflecting the strong discontinuous 
nature of the transition.  
(d) The area fraction of the solid phase to the total area, $A_S/A$, as a function of $\phi$. 
It starts to increase from $\phi_L$ until $\phi_S$ in proportional to $\phi-\phi_L$, indicating the validity 
of the lever rule.  
We also confirmed the similar relation between the number of particles of the solid phase $N_S$ and the total number of particles $N$. 
This means that each coexistence phase retains the same number densities irrespective of $\phi$ in the coexistence region.   
The error bars in (a)-(c) represent the standard deviations of the fittings. For all the data, $\Gamma=3.3$
}
\label{fig:xi_chi}
\end{figure}

As examples of our analysis, here we show details of our fittings of $g_6(r)/g(r)$ for the steel ball system, respectively, in Fig. \ref{fig:supple_xi6}. 
The fittings to $g(r)-1$ are basically the same. 
From these analyses, we obtain the $\phi$-dependence of the spatial correlation length of the hexatic order $\psi_6$, $\xi_6$, 
and that of the translational order, $\xi$, although there are considerable errors due to the small system size. 
The growth of these lengths with an increase in $\phi$ are used to confirm the prediction of the KTHNY theory in a quantitative manner (see Fig. \ref{fig:xi_chi}(a) and (b)). 
We can also see indications that the decays of the envelops of $g_6(r)/g(r)$ and $g(r)-1$ change from exponential to power law, respectively, 
around $\phi_h=0.72$ and $\phi_s=0.735$.

The phase boundaries between liquid, hexatic, and solid phases are determined by standard methods used to study a thermal system, such as 
the nature of the decay of the spatial correlation of bond orientational and translational order (see Fig. \ref{fig:KTHNY}). 
We measured the correlation length of the hexatic order parameter $\psi_6$, $\xi_6$, as a function of the area fraction $\phi$.  
Firstly, the decay of the correlation in the hexatic order parameter changes from 
an exponential to a power-law decay at the liquid-hexatic transition point, $\phi_h$ ($\cong 0.72$); secondly, 
the density correlation changes from an exponential to a power-law decay at the hexatic-solid transition point $\phi_s$ ($\cong 0.735$) 
(see above). 
As shown in Fig. \ref{fig:xi_chi}(a), we observe the steep divergence of $\xi_6$ towards $\phi_h$, which is consistent with the prediction of the KTHNY theory, 
$\xi_6=\xi_6^0 \exp \left[ c'|(1/\phi_h)-(1/\phi)|^{-1/2} \right]$. 
We also fit the relation of $\xi=\xi_0 \exp(c|(1/\phi_S)-(1/\phi)|^{-0.37})$ to the data of $\xi$, although the accuracy of the data is not so high. 
We also observe a sharp peak in the susceptibility of $\psi_6$, $\chi_6=\langle (\psi_6-\langle \psi_6 \rangle)^2 \rangle$ at $\phi_h$ 
(Fig. \ref{fig:xi_chi}(c)). 
In this way, we confirm the presence of a liquid, hexatic, and solid state and determine the phase-transition compositions $\phi_h$ and $\phi_s$.  
However, the estimation of $\phi_s$ may not be so accurate since the range of the analysis of $g(r)-1$ is limited by finite size effects  (see Sec. II B).  
Furthermore, the short-range order developing for $r<4d$ do not allow us to access an asymptotic power law decay of $g(r)-1$ in a wide distance range.  
To avoid finite size effects, we need to perform experiments in a large cell. 
However, we can at least see the very slow algebraic decay of $g(r)-1$ above $\phi_s$.  

We note that, for steel balls, we do not observe any indication of liquid-solid coexistence, also suggesting 
the (nearly) continuous nature of the two transitions. 
The basic phase ordering behavior of a steel ball system 
is thus fully consistent with the KTHNY scenario, as reported previously by Shuttack and his coworkers \cite{shattuck2006crystallization}. 

\subsection{Ordering in a rubber ball system upon densification.}
\begin{figure}[t!]
\begin{center}
\includegraphics[width=8.5cm]{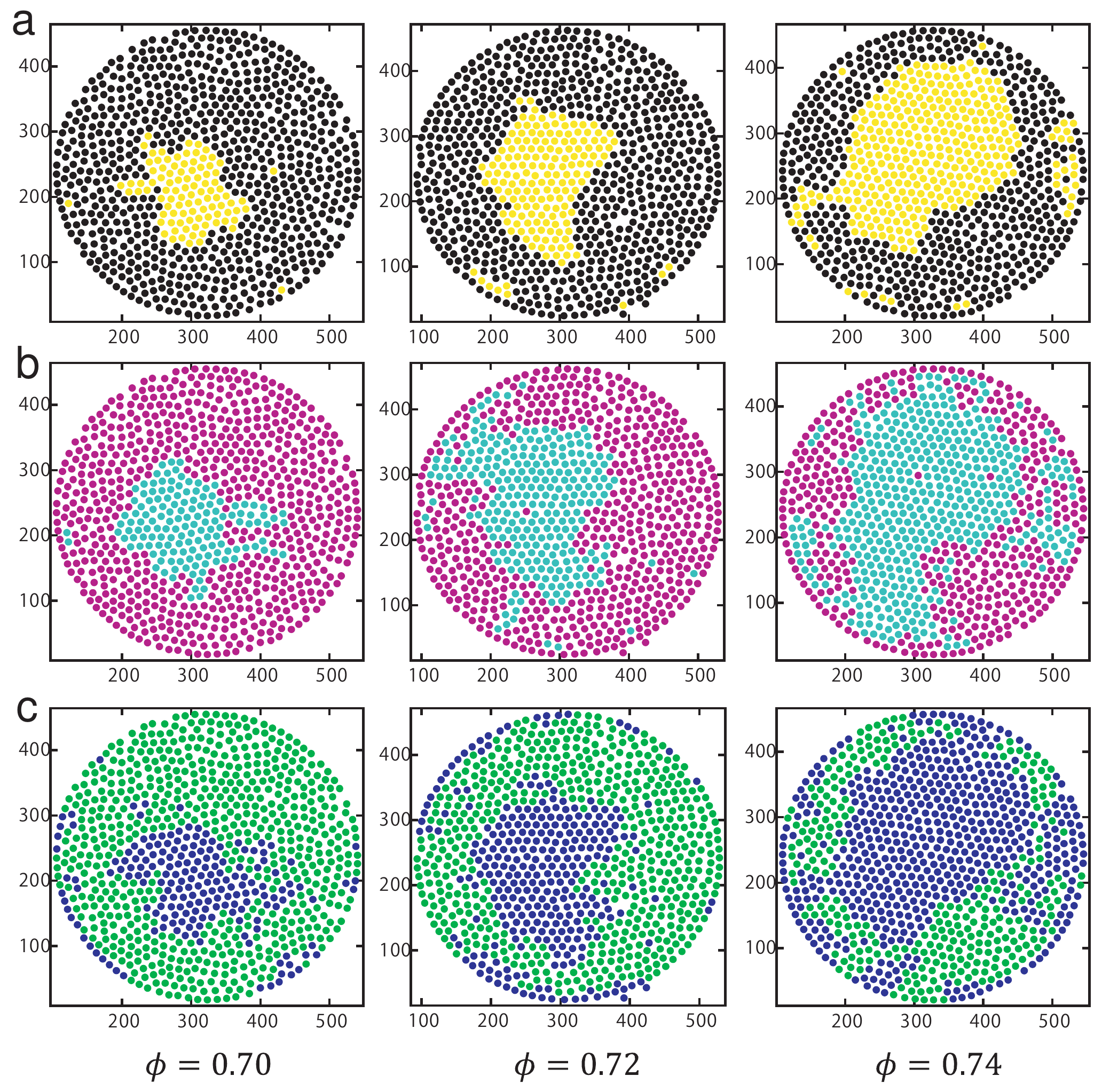}
\end{center}
\caption{Liquid-solid coexistence for rubber ball systems observed at $\phi$=0.70, 0.72, and 0.74, 
which are between $\phi_L$(=0.695) and $\phi_S$(=0.775), as $\Gamma=3.3$.   
The two-phase coexistence can be seen by binarisation by using the following three quantities: 
(a) the hexatic order parameter $\psi_6$ of each particle (here the threshold is chosen as $\psi_6^{\rm th}=0.6$, 
and the pattern is insensitive to the choice for $\phi=0.6-0.7$ (see Fig. \ref{fig:probability})). Yellow (solid) and black (liquid) particles correspond to particles having $\psi_6 \geq \psi_6^{\rm th}$ and 
$\psi_6 <\psi_6^{\rm th}$, respectively.  
(b) the local density $\rho$, which is calculated from the Voronoi area of each particle (the threshold $1/\rho^{\rm th}=9.4$ mm$^2$). 
Blue-green (solid) and pink (liquid) particles correspond to particles having $\rho \geq \rho^{\rm th}$ and 
$\rho<\rho^{\rm th}$, respectively.  
(c) the displacement over 10 s of each particle (the threshold $\delta^{\rm th}$=3.4 mm). 
Blue (solid) and green (liquid) particles correspond to particles having $\delta \leq \delta^{\rm th}$ and 
$\delta>\delta^{\rm th}$, respectively.  
We can see that the solid regions identified by these three quantities are well correlated with each other. 
We can also see the monotonic increase of the solid fraction with an increase in $\phi$.  
}
\label{fig:coexistence}
\end{figure}

\begin{figure}[h!]
\begin{center}
\includegraphics[width=8.5cm]{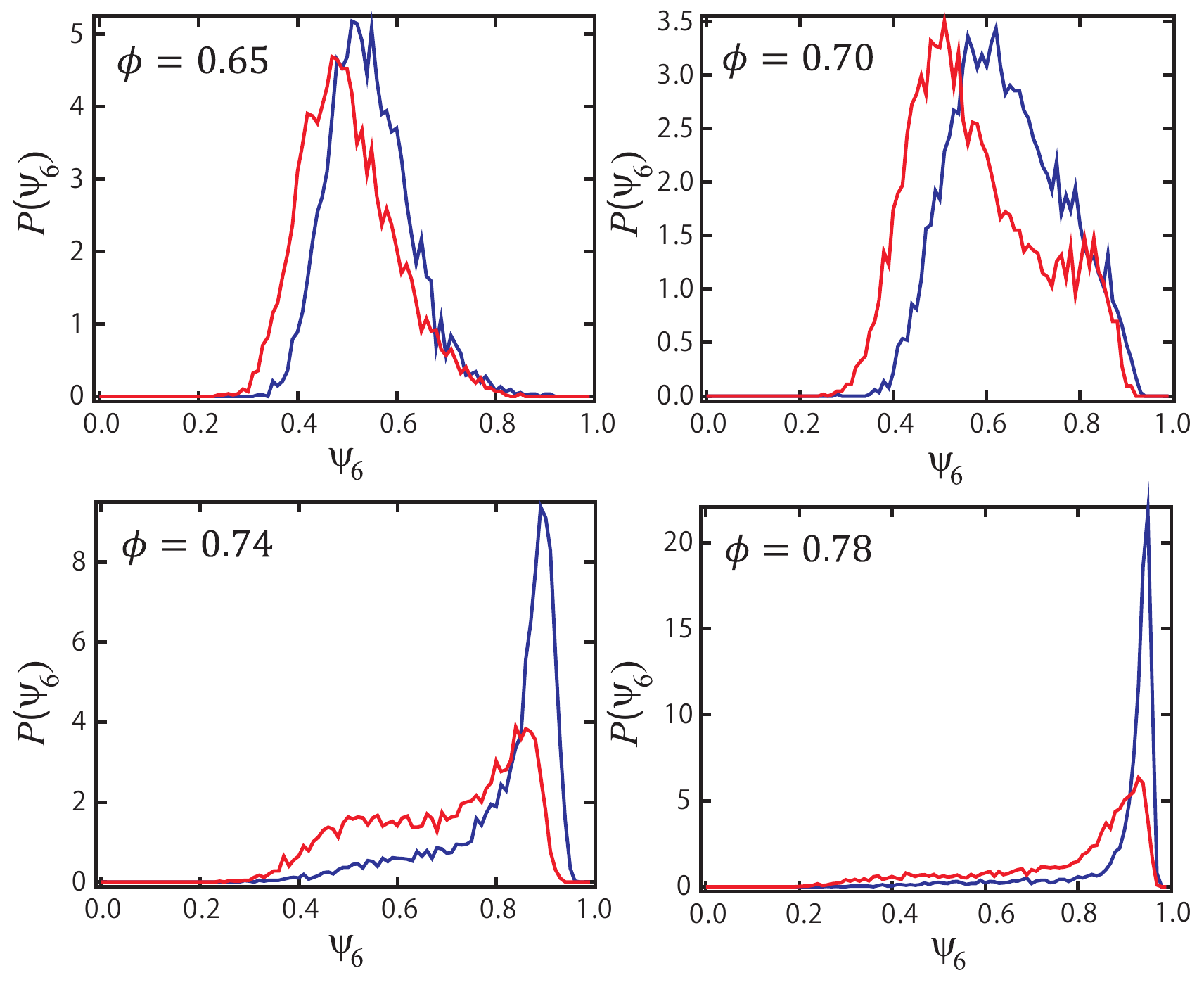}
\end{center}
\caption{
$\phi$-dependence of the probability distribution function of the hexatic order parameter $\psi_6$, $P(\psi_6)$, at $\Gamma=3.3$. 
Blue curves are for a steel ball system, whereas red ones for a rubber ball system.  
For a steel ball system, $P(\psi)$ always has a unimodal shape, whereas for a rubber ball system 
it has a clear bimodal shape for $\phi$ between $\phi_L$ and $\phi_S$. 
The long tails of $P(\psi_6)$ toward low $\psi_6$ in homogeneous solid phases come from defects. 
}
\label{fig:probability}
\vspace{3mm}
\end{figure}

\begin{figure}[h!]
  \begin{center}
　\includegraphics[width =8.cm]{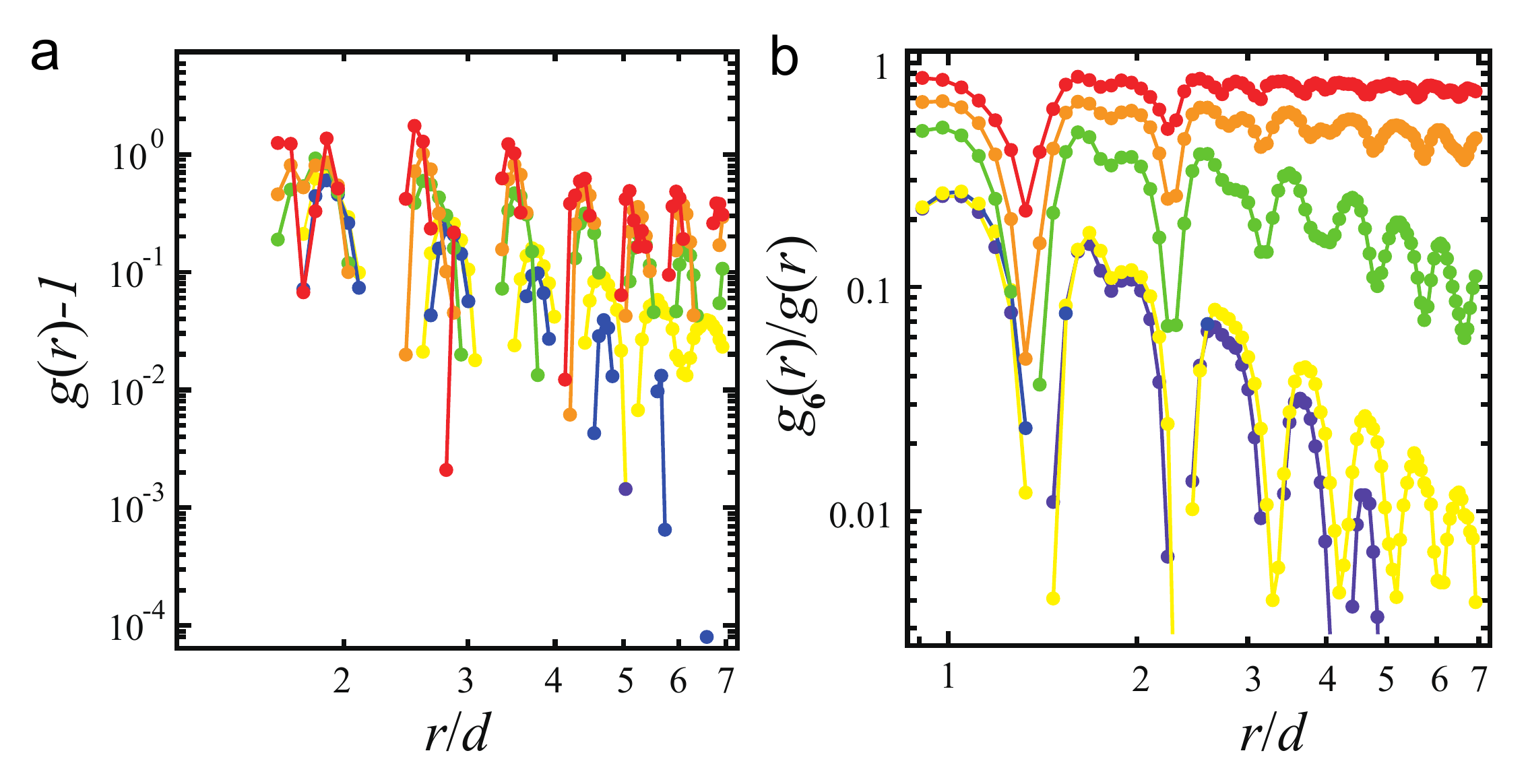}
  \end{center}
  \caption{$\phi$-dependence of $g(r)$ and $g_6(r)/g(r)$ for a rubber ball system at $\Gamma=3.3$.  
    (a) $\phi$-dependence of $g(r)$. The signals are $\phi=$0.65 (bule), 0.69 (yellow), 0.70 (green), 0.74 (orange), and 0.80 (red) 
	from the bottom to the top. (b) $\phi$-dependence of $g_6(r)/g(r)$. 
	The signals are $\phi=$ 0.65 (bule), 0.69 (yellow), 0.70 (green), 0.74 (orange), and 0.80 (red) from the bottom to the top.  
	}
  \label{fig:rubber_g(r)}
\end{figure}

Next, we focus on the liquid-solid transition in a rubber ball system. 
Unlike in the above steel ball system, we observe in this system the two-phase coexistence of the liquid and solid phases separated by a rather sharp interface 
at a certain range of $\phi$ (see Fig. \ref{fig:coexistence}).  
The two-phase coexistence can also be clearly seen from the bimodal shape of the probability distribution function of 
the hexatic order parameter $P(\psi_6)$, as shown in Fig. \ref{fig:probability}. 
Below $\phi_L$, the spatial correlations of hexatic and translational order are both short-range and decayed nearly exponentially,  
as shown in Fig. \ref{fig:rubber_g(r)}.  
As can be seen in Fig. \ref{fig:xi_chi}(b), neither the translational nor orientational correlation length, 
$\xi$ and $\xi_6$, exhibit any growth with an increase in $\phi$, unlike the case of a steel ball system. 
There is also no increase in the susceptibility around the phase transition points, as shown in Fig. \ref{fig:xi_chi}(c). 
Since the analysis of the decay of the spatial correlation functions is not useful in the coexistence region,  
we distinguish a liquid state, a solid state, and a coexistence state 
on the basis of measurements of the local orientational order, the local density (or, area fraction), and the local mobility, which has a link to the granular temperature, 
(see Fig. \ref{fig:coexistence}). 
We can see these three different quantities including both static and dynamical ones can identify the two-phase coexistence 
in a consistent manner, although the analysis of an instantaneous structure leads to some inaccuracy due to short-time fluctuations of the interface 
(see below on a possible origin). 
We can also clearly see that the fraction of the solid phase monotonically increases with an increase in $\phi$. 
The area fraction of the solid phase determined from the hexatic order parameter is shown in Fig. \ref{fig:xi_chi}(d), which indicates 
the validity of the lever rule (see below). 
A discontinuous first-order-like phase transition often accompanies hysteresis and metastability. 
We made all the observation after attaining a steady state (typically after 10 min from the initiation of vibrational driving) and the behavior was very reproducible.   
We did not observe any indication of hysteresis mainly because measurements are always done after a steady state is reached 
and there is no way to change the area fraction continuously.  
So we identify $\phi_L$ and $\phi_S$ as the lower and upper boundary of the solid and liquid phase, respectively. 

\begin{figure}[h!]
  \begin{center}
   \includegraphics[width =8.cm]{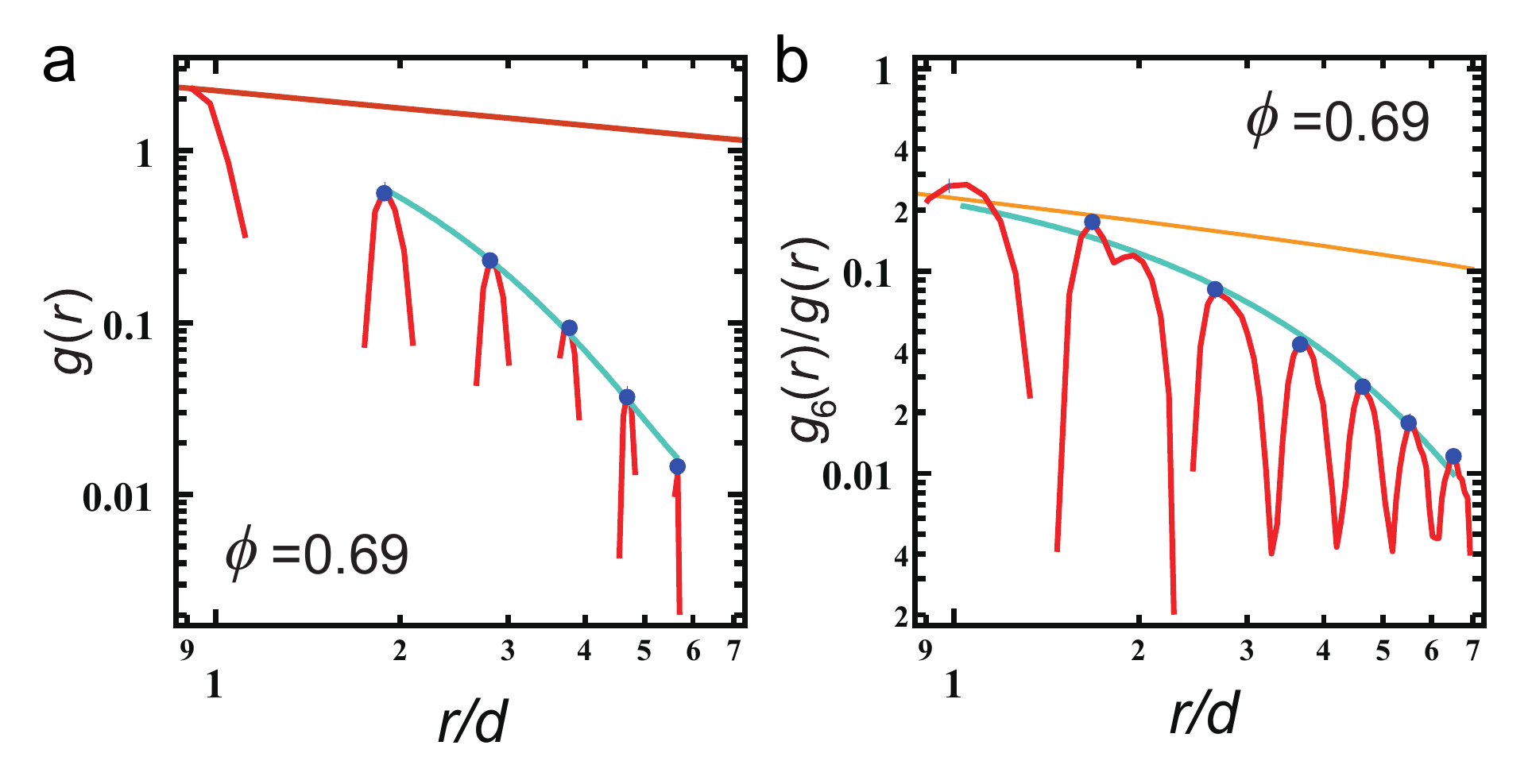}
  \end{center}
  \caption{Examples of the analysis of the spatial decay of $g(r)$ and $g_6(r)/g(r)$ for a rubber ball system at $\Gamma=3.3$.
Here we show the analysis of $g(r)$ (a) and $g_6(r)/g(r)$ (b) at $\phi=0.69$, which is very close to $\phi_L=0.695$.   
The blue points are used for the fittings. 
The blue curves are exponential fittings, whereas the red and orange lines 
have a slope of $-1/3$ and $-1/4$ respectively.  
In a one phase region, the correlation of the bond orientational order decays exponentially and 
there is no indication of a power-law decay even near $\phi_L$. 
The extracted correlation lengths in this manner are plotted as a function of $\phi$ in Fig. \ref{fig:xi_chi}(b).  
We note that in the two-phase coexistence region, 
this type of analysis is not meaningful.  
}
  \label{fig:rubber}
\end{figure}

Here we show examples of of our fittings of $g(r)$ and $g_6(r)/g(r)$ for the rubber ball system at $\phi=0.69$ in Fig. \ref{fig:rubber}.  
We find no systematic $\phi$-dependences for $g(r)$ and $g_6(r)/g(r)$ (see Fig. \ref{fig:xi_chi}(b)). 
We can see in Fig. \ref{fig:rubber} that even at $\phi=0.69$, which is very near $\phi_L=0.695$, 
the decays of density correlation and hexatic order correlation are both exponential and much faster than the power law decays of exponent $-1/3$ and $-1/4$, 
respectively.  

Above $\phi_S$, we see a homogeneous solid phase. 
For $\phi_L<\phi<\phi_S$, we observe the coexistence of the liquid and the solid phases. 
We confirm that the former has the upper bound $\phi$ of the liquid phase, $\phi_L$, and the latter has the lower bound $\phi$ of the solid phase, $\phi_S$. 
We determine $\phi_L=0.695$ and $\phi_S=0.775$.  
The area fraction of each phase obeys the lever rule in the coexistence region, as shown in Fig. \ref{fig:xi_chi}(d).

The strong discontinuous nature of the transition is also confirmed by that fact that 
there is neither the divergence of $\xi_6$ (Fig. \ref{fig:xi_chi}(b)) nor  the sharp increase in $\chi_6$ (Fig. \ref{fig:xi_chi}(c)) 
at the phase boundary.  
We also show the distribution function of $\psi_6$, $P(\psi_6)$, in Fig. \ref{fig:coexistence}(e) and (f). 
We can see a clear bimodal distribution for the coexistence region of $\phi$, which is another clear indication of the liquid-solid coexistence. 
Such a signature of the strongly first-order transition is absent for a system of steel balls. 

We find that the coexistence can be seen not only by the local packing symmetry characterised by the hexatic order parameter (Fig. \ref{fig:coexistence}(a)), 
but also by the area fraction $\phi$ (Fig. \ref{fig:coexistence}(b)) and the in-plane displacement amplitude (Fig. \ref{fig:coexistence}(c)).   
The last point indicates that the effective (granular) temperature $T^\ast$ defined by the kinetic energy is spatially inhomogeneous 
for the liquid-solid coexisting state. Thus, the solidity is linked not only to the static quantities such as the area fraction and the 
bond orientational and translational order, but also to the effective temperature. 
This is a very unique feature of this non-equilibrium steady state, which is maintained by continuous 
vibrational energy input and dissipation due to inelastic interparticle collisions. 
This phenomenon has some similarity to the inhomogeneization of a granular gas due to inelastic collisions \cite{jaeger1996granular,aranson2006patterns,goldhirsch1993clustering,puglisi1998clustering}. 
Coexistence of phases with different granular temperatures was also reported for a quasi-2D system where bilayer formation is allowed \cite{urbach2004,urbach2005dynamics,urbach2008effect,urbach2009effects,luu2013capillarylike,clerc2008liquid}, however, 
the underlying mechanism may be quite different, as will be discussed later. 
The link between the solidity and the effective temperature indicates that the interfacial profile may be 
related to the spatial gradient of not only the area fraction and the hexatic order but also the temperature (i.e., the kinetic energy). 
This possible dependence of the interfacial profile on the mobility, or the effective temperature, is unique to non-equilibrium open systems 
and absent in thermal equilibrium systems (see, e.g., \cite{brey1998hydrodynamics}) .

\subsection{State diagrams.}
\begin{figure}[h!]
\begin{center}
\includegraphics[width=8.0cm]{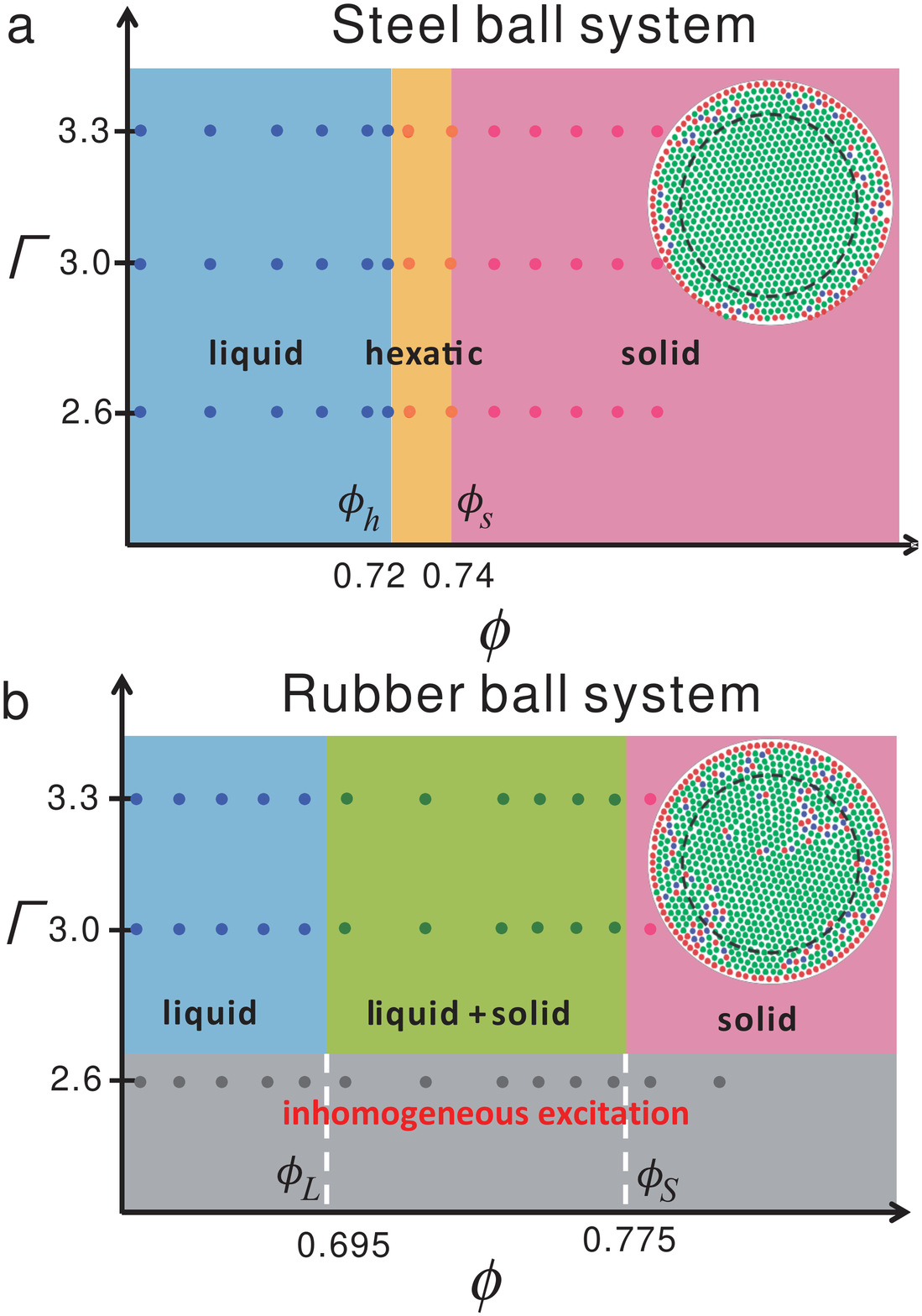}
\end{center}
\caption{Phase diagram of driven monolayer sphere systems. 
Filled circles represent state points where the measurements were made (blue: disordered liquid; orange: hexatic phase; pink: solid; 
green: a coexistence of liquid and solid; gray: no well-defined stationary state due to inhomogeneous excitation). 
(a) Steel ball systems, which obey the KTHNY-like scenario. In this case, the state behavior does not depend 
on $\Gamma$ in the range studied. 
(b) Rubber ball systems, which show the distinct discontinuous transition between liquid and solid states. 
In this case, the dynamical steady state is realized only above $\Gamma \cong 2.7$. 
This higher threshold value of $\Gamma$ is presumably due to the inelastic nature of collisions of rubber balls with the confining plates.  
The images in (a) and (b) are configurations in the solid phase at $\phi=0.78$ for steel and rubber ball systems, 
respectively. Green, blue, and red particles have local hexagonal, pentagonal, and heptagonal structures, respectively. 
The inner part surrounded by the red dashed circle has less order (i.e., more defects) for the rubber ball system than for the steel one.    
}
\label{fig:PD}
\end{figure}

On the basis of these results, we draw the phase (more strictly, state) diagrams for steel and rubber balls, which are shown in Fig. \ref{fig:PD}. 
Here we include the dependence of the phase behavior on $\Gamma$. 
In general, we need $\Gamma$ larger than a critical value $\Gamma_c$ to maintain a dynamical steady state. 
Below $\Gamma_c$, the energy injection becomes inhomogeneous. The gray region labeled ``inhomogeneous excitation" in Fig. \ref{fig:PD}(b) 
is in such an inhomogeneous state. 
This is because a well-defined dynamical steady state can be maintained only when the energy injected by vertical vibration 
overcomes the effects of the gravity and the energy loss due to inelasticity of collisions of balls and the walls \cite{olafsen1998clustering,nie2000dynamics,khain2011hydrodynamics} 
(see also Appendix A-2 and  Appendix B). 
We find the critical value of $\Gamma$ to maintain a dynamical steady state is higher 
for the rubber ball system ($\Gamma_c \sim 2.7$) than for the steel ball system  ($\Gamma_c \sim 1.3$), 
reflecting the stronger inelastic nature of collisions of the former with the confining plates.  
We note that the transition area fractions, $\phi_h$, $\phi_s$, $\phi_L$, and $\phi_S$, are all independent 
of $\Gamma$ within the accuracy of our measurements. 

The insets of Fig. \ref{fig:PD}(a) and (b) show the local orientational order in the solid phases formed in a steel and rubber ball system at $\phi=0.78$, respectively. 
We can clearly notice that the amount of defects is much larger for a rubber ball system than for a steel one. 
This can also be seen in the shape of $P(\psi_6)$, which has a larger tail toward low $\psi_6$ for a rubber ball system than for a steel one. 
This may be just a consequence of the difference in the lower stability limit $\phi$ of the solid phase between a steel and a rubber ball system, but 
there might be other fundamental reasons. 

\subsection{Dissipation-induced wetting.}
\begin{figure}[h!]
\begin{center}
\includegraphics[width=8.cm]{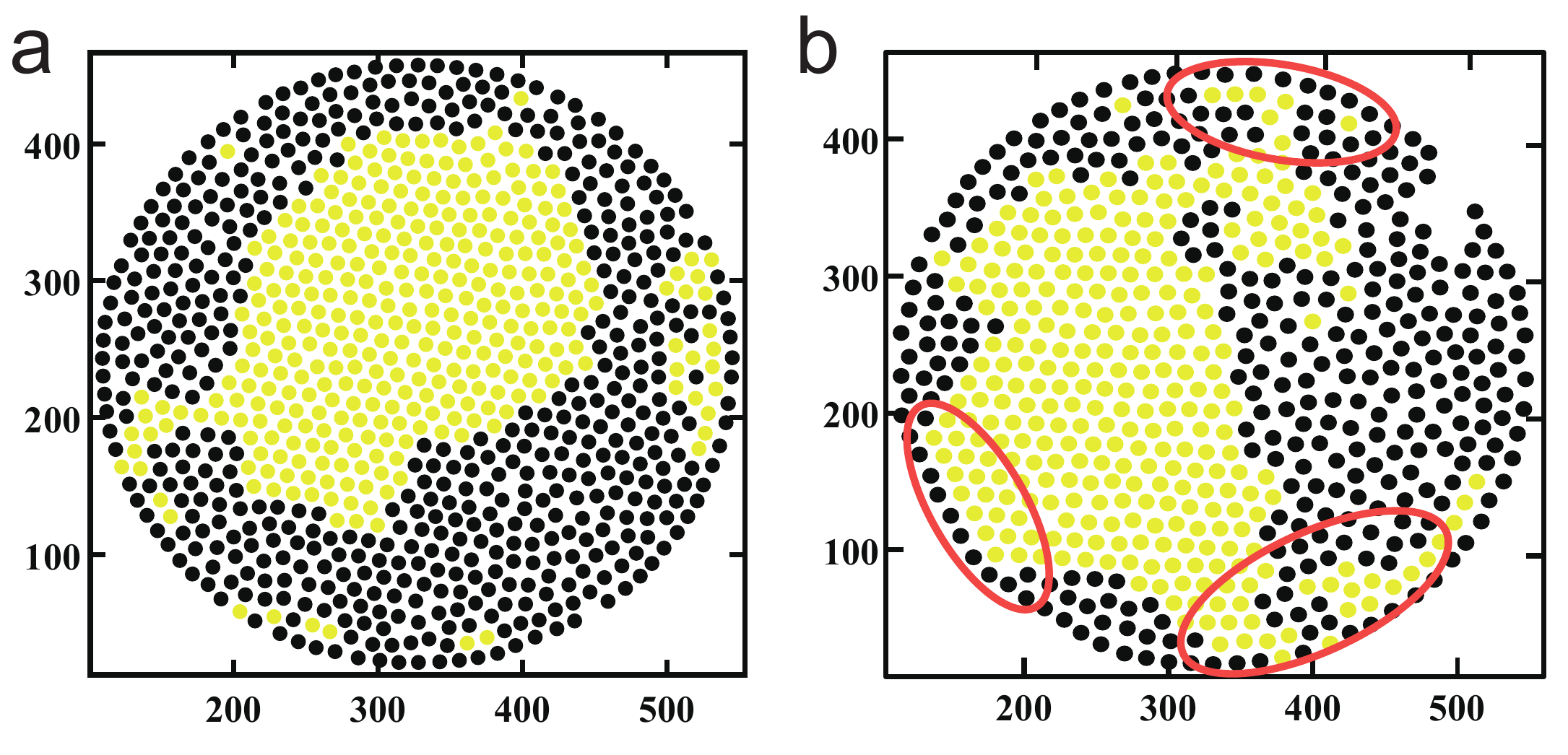}
\end{center}
\caption{Wall effects on the liquid-solid coexistence.
(a) A coexistence of the liquid and solid phase confined by a circular steel wall for $\phi=0.74$. 
The liquid phase preferentially wets the wall. 
(b) The same as (a) for a softer wall whose surface is covered by a silicone rubber film. 
In this case, the solid phase partially wets the wall (see the regions surrounded by the red lines).
}
\label{fig:boundary}
\end{figure}

We also note that the solid phase is always formed in the middle part of the container far from the side wall (see Fig. \ref{fig:coexistence} 
and Fig. \ref{fig:boundary}(a)).  
This may be explained by a larger restitution coefficient of particle-side wall collision than particle-particle one. 
The hard side wall prefers the liquid phase with high $T^\ast$. 
We confirm that a softer side wall (covered by a silicone rubber film) is statistically more wettable to the solid phase (Fig. \ref{fig:boundary}(b)), although 
wetting is rather modest likely because the curved wall inevitably induces elastic distortion of 
the solid phase. Thus, this wetting phenomenon may be regarded as \emph{dissipation-induced wetting}.

\subsection{Inelasticity-induced demixing.}
\begin{figure}
\begin{center}
\includegraphics[width=8.cm]{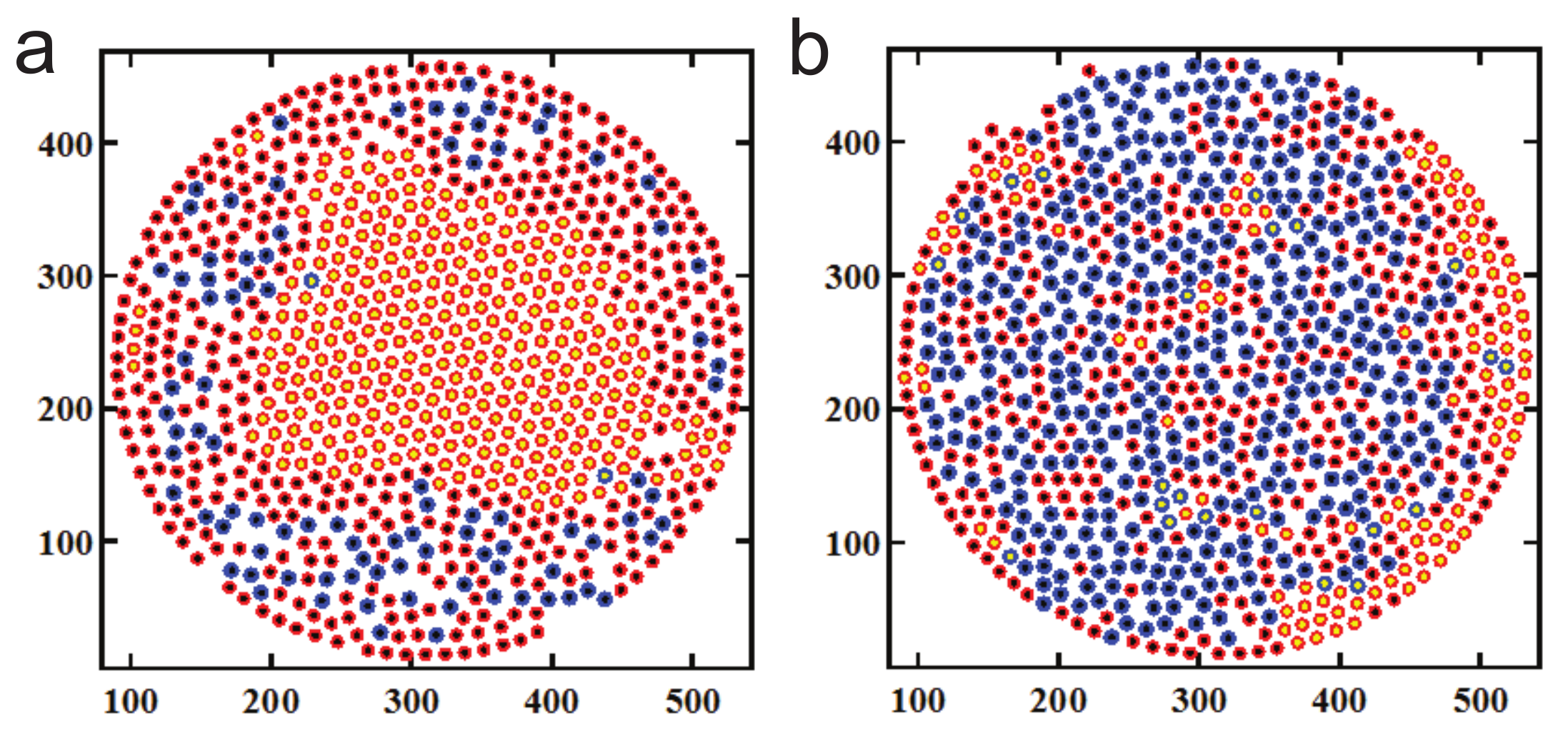}
\end{center}
\caption{Phase behavior of mixtures of rubber and steel balls.
(a) Snapshot of a 1:7 mixture of steel and rubber balls (the total area fraction of $\phi=0.72$) driven at $\Gamma=3.3$. 
Here we can see a coexistence of the solid phase purely made of rubber balls and the liquid phase which is a mixture 
of steel and rubber balls. 
(b) Snapshot of a 1:1 mixture of steel and rubber balls (the total area fraction of $\phi=0.72$) driven at $\Gamma=3.3$. 
The color of the outer shell of each particle represents the type of the particle (blue: steel ball; red: rubber ball), whereas 
that of the inner core the degree of the hexatic order (black: low $\psi_6$; yellow: high $\psi_6$). 
}
\label{fig:mixture}
\end{figure}

Finally, we briefly mention a related interesting 
phenomenon we observe in a mixture of steel and rubber balls. 
There are some studies on inelasticity-induced demixing \cite{brey2005energy,serero2006hydrodynamics,brito2008segregation,garzo2009segregation,brito2009competition,windows2014inelasticity};  
however, there is no example associated with a liquid-solid transition.   
Here we report inelasticity-induced demixing, which is linked to the formation of an ordered solid phase.  
When we mix a small amount of steel balls with rubber balls, we observe steel balls are 
completely expelled from the solid region and included in the liquid region (see Fig. \ref{fig:mixture}(a)). 
We note that the solid phase is made of only rubber balls: \emph{dissipation-induced demixing}. 
Here we stress that in such a mixture the energy input rate to steel balls is higher than that to rubber balls. 
On noting that steel balls have a larger kinetic energy due to its larger restitution coefficient, 
the above argument naturally explains the preferential inclusion of steel balls into the liquid phase. 
If the fraction of steel balls is too large, the coexistence conditions cannot be 
satisfied simultaneously and thus 
the formation of the macroscopic ordered solid phase is largely prohibited even at the same area fraction $\phi$. 
As a result, a rather random mixed liquid state is realized (see Fig. \ref{fig:mixture}(b)). 

\section{Discussion}

\subsection{Comparison of our study with previous works on driven quasi-2D systems.}

Here we consider the relationship of our results with those of pioneering works on phase orderings in quasi-2D driven granular systems \cite{urbach2004,urbach2005dynamics,urbach2008effect,urbach2009effects,luu2013capillarylike,clerc2008liquid}, in which 
bilayer formation is allowed. 
We note that there is a crucial difference in the dependence of the phase behavior on the strength of inelasticity:  
For systems, where a bilayer can be formed, ordering (towards bilayer solid) is suppressed by inelasticity \cite{urbach2008effect,urbach2009effects,clerc2008liquid}. 
We emphasize that this is completely the opposite to our case:  inelasticity helps ordering (towards monolayer solid). 
See Appendix A-1 for a more detailed discussion including 
an intuitive explanation on the cause of the difference.    
We also note that the monolayer-bilayer formation can take place as a function of $\Gamma$ 
\cite{urbach2004,castillo2012fluctuations}, 
whereas our transition cannot be induced by changing $\Gamma$ as we can see from the phase diagrams in Fig. \ref{fig:PD}. 
Near $\Gamma_c$, we see a crossover from inhomogeneous excitation to homogeneous one, but this transition 
as a function of $\Gamma$ has an origin essentially different from the transition as a function of $\phi$ (see Appendix A-2 and B for the details).
The above-mentioned fundamental differences between our work and the previous studies strongly indicate that the underlying physics is 
essentially different between the two cases even on a qualitative level. 
In Appendix A-3, we also mention another type of inhomogeneization in driven inelastic particles, gas-liquid coexistence \cite{roeller2011liquid}. 

Finally, we mention the 2D melting of a solid phase with an increase in $\Gamma$ \cite{olafsen2005two}. 
As described in the introduction, this melting is induced by the increase in the defect density, which is 
caused by the increase in hight fluctuations of particles. Since our 2D constraint $h/d \sim 1.3$ is much stronger than the one used in this work 
($h/d=1.6$),  we did not see any indication of $\Gamma$-induced melting. 
It is interesting to study in the future how and at which $h/d$ the behavior changes.   
   
\subsection{Roles of inelastisity and friction in phase-ordering behavior.}
First we qualitatively consider what is the origin of the difference in the phase-transition behavior between the steel and rubber ball systems. 
Steel and rubber balls differ not only in the inelasticity and friction upon interparticle collisions but also in the softness of interparticle interactions. 
It is rather well established that the KTHNY-like behavior, or a transition from 
liquid to solid via the intermediate hexatic phase, is very robust for 2D `thermal' systems, irrespective of the types of interparticle interactions:  
Essentially the same behavior was observed not only for hard disks but also for particles interacting with soft repulsive potentials 
and attractive (e.g., Lennard-Jones) potentials  (see, e.g., Refs. \cite{keim2010melting,wang2011two,mazars2013melting}),    
although it has recently been shown that the liquid-hexatic transition depends delicately on the softness of the interaction \cite{kapfer2014soft}. 
This robustness of the KTHNY-like behavior irrespective of the nature of interactions 
is a natural consequence of the fact that hexatic order is formed primarily by geometrical packing effects:  
When hard-sphere-like particles are packed, this local configuration is entropically favored. 
This suggests that the nature of the liquid-solid transition between the steel and rubber ball systems is not controlled by 
the softness of particles, but by the dissipation due to the inelasticity of particles and the friction.

Now we consider how inelasticity and friction can change the nature of the liquid-solid transition. 
Steel balls are not perfectly elastic but the system still behaves like a thermal system. On the other hand, the behavior of the rubber ball system is distinctly different from the behavior 
of its thermal counterpart. So the question can be rephrased as what physical mechanism controls the transition 
from apparently thermal to athermal behavior  
when we change the degree of inelasticity and friction from steel to rubber balls. 
For a vertically vibrated monolayer system, the particle-(top and bottom) wall collision frequency, $f_{pw}$, is higher than the particle-particle collision frequency, 
$f_{pp}$, as long as $\Gamma >\Gamma_c$. 
We estimated $f_{pw}$ as roughly 100 Hz by illuminating light from a low angle from the horizontal plane and 
$f_{pp}$ as roughly 20-50 Hz, depending upon on the local $\phi$, by using normal illumination from the top, with a fast camera. 
For an inelastic particle system to apparently behave like an elastic thermal system, the energy dissipated by each particle-particle collision should be fully recovered by energy input through  
particle-wall collisions before the next collision with a particle takes place. 
As long as this condition is satisfied, even a dissipative system apparently behaves like a thermal system, as our steel ball system does. 

The athermal nature is accompanied by inhomogeneization of a system and the resulting coexistence of 
two phases with different effective granular temperatures. 
Such two-phase coexistence characteristic to athermal systems was already discovered for a quasi-2D driven granular system, in which  
bilayer formation is allowed unlike our strictly monolayer system \cite{urbach2004,urbach2005dynamics,urbach2008effect,urbach2009effects,luu2013capillarylike,clerc2008liquid}. 
In such a case, the different granular temperatures can be explained by 
the difference between monolayer excitation and bilayer excitation (e.g., differences in effective mass and type of excitation).
However, this mechanism cannot explain the two-phase coexistence we found in a driven monolayer system. 
Below we consider the origin of inhomogeneization and the resulting coexistence of two phases with different granular temperatures on a qualitative level.  
 
Here we consider a gedanken experiment, where the inelasticity is gradually increased in a continuous manner.  
With an increase in the inelasticity of particles, the perfect recovery before the next interparticle collision becomes more and more difficult. 
With an increase in the particle density, or $\phi$, the inter-particle collision frequency $f_{pp}$, which controls the rate of energy dissipation,  increases but 
with a rather constant particle-wall collision frequency $f_{pw}$, which makes this recovery process less efficient. 
The energy dissipation is due to both the inelastic and the frictional nature of interparticle collisions. 
We believe that this crossover from the perfect to imperfect recovery of the kinetic energy during $1/f_{pp}$ with an increase in the particle density 
destabilizes a non-equilibrium steady state with spatially homogeneous density and granular temperature, leading to the transition from an apparently thermal 
to strongly athermal behavior. 
Now we qualitatively consider how the imperfect recovery makes a system inhomogeneous: In such a situation, the degree of recovery of the kinetic energy 
of a particle should decrease with an increase in local $\phi$ since $f_{pp}$ increases with an increase in $\phi$. Thus, higher density regions formed 
by spontaneous fluctuations should transiently have lower kinetic energy than lower density regions, which leads to lower pressure in the former. 
This mechanism is essentially the same as that of clustering instability in dissipative gases \cite{goldhirsch1993clustering}.  
So the higher density regions are further compressed, resulting in the enhancement of density fluctuations.  
However, an increase in density eventually causes the increase in the inter-particle collision frequency and thus 
the increase in pressure.  When the horizontal pressure is spatially homogeneized, this development of density fluctuations stops, ending 
in two-phase coexistence. In this way the system finally reaches a steady state, where the energy input rate and the dissipation rate is balanced 
while keeping two phases with different $\phi$ and granular temperature $T^\ast$. 
Thus, a high enough density region and a low enough density region should coexist in a steady state for a certain range of the average $\phi$. 
The average horizontal pressure in the two phases should be the same since the mechanical force balance is to be satisfied across the domain interface. 
Because of its mechanical nature, this equal pressure condition may be hold even in an athermal condition (see below, however, on the 
effect of interparticle friction). However, the fluctuations of pressure may be much larger than in a thermal system, since the particle velocity 
is determined not only by interparticle collisions but also by particle-wall collisions.  
The another constraint comes from the condition for a steady state: the balance between the energy input rate and the dissipation rate should be balanced in each phase.  
The low density liquid phase is characterized by low $f_{pp}$, which means low energy dissipation and results in high $T^\ast$, whereas the high density solid 
phase by high $f_{pp}$, which results in low $T^\ast$.  
This situation is realized under a homogeneous energy input, or spatially homogeneous $f_{pw}$. 
On the basis of this physical picture, below we consider a principle behind the two-phase coexistence on a phenomenological level. 

Here we note that it is interesting to study the dependence of the phase-transition behavior as a function of inelasticity. 
However, a precise control of particle inelasticity is not easy in experiments, and thus numerical simulations may be more promising.  

\subsection{How to approach the problem}
Before discussing a principle behind the two-phase coexisting more theoretically, here we consider which theoretical framework 
is suitable for the description of what we discussed above.   
One candidate is a hydrodynamic theory for a confined granular system, which was recently developed by Brito et al. \cite{brito2013hydrodynamic}. 
It was shown that under a situation where the thermostat does not inject momentum but only energy, the equations for the conserved density field and momentum density are the continuum equation and the Navier-Stokes one, as in the case of usual fluids, and the equation for the non-conserved (granular)  temperature field is a balance equation for the energy. A collisional model for a 2D system was also developed on the same basis (see Sec. IV of Ref. \cite{brito2013hydrodynamic}); however, in this model the stationary temperature is density independent and the pressure increases monotonically with density, and accordingly there is no two-phase coexistence, contrary to our experimental observation.  In this model, the granular temperature in a homogeneous stationary state was assumed to be determined by a balance between energy dissipation and injection \cite{brito2013hydrodynamic}, but the energy dissipation due to an effective friction proportional to a particle velocity was not considered. Although it is interesting to take this effect into account in this theory and to study whether it can induce two-phase coexistence, here we take a more phenomenological approach. 

To take this velocity-dependent dissipative force into account in a natural manner, we consider a thermal condition to maintain a steady state in a quasi-2D system, on the basis of a Langevin-like equation of motion (see, e.g., \cite{sarracino2010granular}). 

\subsection{A physical principle behind two-phase coexistence in a non-equilibrium steady state. }
Now we consider a principle behind two-phase coexistence in a non-equilibrium steady state.  
For a thermodynamic system, in order for two phases of the same substance in
contact to be in equilibrium, there must be mechanical, thermal, and chemical equilibrium, i.e.,  
(i) equal pressure, (ii) equal temperature, and (iii) equal chemical potential. 
For our driven granular system, however, condition (ii) is not be applicable because of the out-of-equilibrium nature. 
On the basis of a physical picture we described in Sec. IV.B, here we focus our attention on this condition, using a model of granular Brownian motion \cite{sarracino2010granular}.

The change in particle velocity due to a binary instantaneous collision between particle $i$ and $j$ with the same mass $m$ is given by 
$\mathbf{v}_i=\mathbf{v}'_i-\frac{1+\alpha}{2}\left[(\mathbf{v}'_i-\mathbf{v}'_j) \cdot \hat{\mathbf{n}} \right] \hat{\mathbf{n}}$, 
$\mathbf{v}_j=\mathbf{v}'_j+\frac{1+\alpha}{2}\left[(\mathbf{v}'_i-\mathbf{v}'_j) \cdot \hat{\mathbf{n}} \right] \hat{\mathbf{n}}$, 
where $\mathbf{v}$ and $\mathbf{v}'$ are the velocity after and before the collision respectively, $\hat{\mathbf{n}}$ is the unit vector 
jointing the centres of particles and $\alpha$ is the restitution coefficient ($0 \leq \alpha \leq 1$) which is equal to 1 in the elastic case. 
In order to maintain a steady state, we need an external energy source that is coupled to every particle in the form of a thermal bath. 
In our case, this external energy is supplied by collisions with vibrating walls confining a granular monolayer. 
The motion of a particle $i$ is then described by the following stochastic equation \cite{sarracino2010granular}: 
\begin{eqnarray}
m \frac{d \mathbf{v}_i(t)}{dt}=\mathbf{f}_i(t)-\gamma \mathbf{v}_i(t) +\mbox{\boldmath$\zeta$}_{b}(t).  \label{eq:BD}
\end{eqnarray}  
Here $\mathbf{f}_i$ is the force taking into account the collisions with other particles, $\gamma=m/\tau$ is a drag coefficient characterizing the velocity decay 
towards a steady state, whose timescale we express by $\tau$,  
and $\mbox{\boldmath$\zeta$}_{b}(t)$ is random white force noise exerted by particle-wall collisions, with $\langle \mbox{\boldmath$\zeta$}_{b}(t) \rangle=0$ and 
$\langle \zeta_{b,i\alpha}(t) \zeta_{b,j\beta}(t')\rangle=2T_{b}\gamma \delta_{ij}\delta_{\alpha \beta} \delta(t-t')$ 
($T_{b}$ being the effective bath temperature). 
Here we note that the viscous term in Eq. (\ref{eq:BD}) takes into account the friction among particles and energy transfers 
between various degrees of freedom (e.g, the friction between particles and the walls) \cite{puglisi1998clustering}. 

Here we mention the random nature of force noise. 
In our system, the excitation itself has a well-defined frequency, but the randomization of the energy injection by interparticle collision 
and the wall roughness leads to a self generated effective white bath in a long timescale. 
The Langevin-type approach can be rationalized by the fact that Brownian dynamics simulations 
can fully reproduce the behavior of a driven monolayer granular system of weakly polydisperse steel balls even without any adjustable parameters \cite{watanabe2011}. 
Its validity to a rubber ball system is not obvious, but our visual inspection of the motion of steel and rubber balls at least 
indicates the same type of random motion. Although the exact nature should be investigated, e.g., by measuring the velocity distribution function in the future, 
this visual inspection and the consideration in Sec. VI.C tell us that in a steady state 
the physics may be the same between the two systems at least on a phenomenological level. So we assume that the randomization 
also takes place efficiently for a rubber ball system.    

A stationary state is maintained since the effect of the external energy source balances the energy lost by interparticle collisions and  particle-wall collisions.  
The key parameters of the system is the characteristic velocity decay time, $\tau$, and the packing fraction $\phi$. 
Multiplying the above stochastic equation Eq. (\ref{eq:BD}) by $\mathbf{v}$ and averaging yields 
\begin{eqnarray}
m \frac{d \langle \mathbf{v}(t)^2 \rangle}{dt}=\langle \mathbf{v}(t) \cdot \mathbf{f}(t) \rangle- \gamma \langle \mathbf{v}(t)^2 \rangle +\langle \mathbf{v}(t) \cdot \mbox{\boldmath$\zeta$}_{b}(t) \rangle.  
\label{eq:BD2}
\end{eqnarray} 

Here we note that the granular temperature $T^\ast$ is defined as $T^\ast=\frac{1}{2}m \langle \mathbf{v}^2 \rangle$ 
and thus the left-hand side of the above equation can be written as $2 \partial T^\ast/\partial t$. 
The first term in the right-hand side of Eq. (\ref{eq:BD2}) can be written as 
$\langle \mathbf{v}(t) \cdot \mathbf{f}(t) \rangle=-\langle \Delta E \rangle_{col}$, where 
$\Delta E=\frac{1}{8m}(1-\alpha^2) \left[(\mathbf{v}_1-\mathbf{v}_2) \cdot \hat{\mathbf{n}} \right]^2$, represents the average energy dissipation rate due to interparticle collisions. 
$\langle \cdots \rangle_{col}$ is the collision average, whose expression is known for granular gas \cite{sarracino2010granular} but not for high-density liquid or solid 
because of the difficulty associated with not only excluded volume effects but also recollisions and memory effects. 
We express this term $\langle \mathbf{v}(t) \cdot \mathbf{f}(t) \rangle=-\langle \Delta E \rangle_{col}$ as $2 \mathcal{D}$.
Although its exact form is not clear, it should be a decreasing function of $\alpha$ and an increasing function of $\phi$, and $T^\ast$.  
The second term is rewritten from the definition of $T^\ast$ as $2 T^\ast/\tau$. 
Finally, the third term is the total energy input rate and written as $\langle \mathbf{v}(t) \cdot \mbox{\boldmath$\zeta$}_{b}(t) \rangle =2 \gamma T_{b}/m=2T_{b}/\tau$.  

Here we briefly consider the role of interparticle friction. Although the friction coefficient itself may be similar between the steel and rubber balls, the rubber ball may have a larger coefficient of rolling friction with the walls as well as another ball upon collision 
than the steel ball does. On a phenomenological level, this effect can be included in the coefficient $\gamma$; then,  $\gamma$, or $\tau$, should be an increasing function of $\phi$.  
The friction also affects the term $\mathcal{D}$ (see, e.g., Ref. \cite{risso2005friction}).   
We also note that the presence of interparticle friction also induces forces tangential to the domain interface. This may perturb the interface position 
and induce large fluctuations of the interface. Such a signature can indeed be seen in Fig. \ref{fig:coexistence}, although it might be largely due to the large pressure fluctuations 
unique to a driven granular system (see Sec. IV.B). 
For simplicity, however, we do not consider this dynamical effect when we discuss the phase coexistence. 

Although the difference in the friction property between steel and rubber balls may play a role in the observed difference 
in the phase transition behavior, the coefficient of rolling friction is generally small for smooth spherical particles.  
Thus, we argue that the difference in the restitution coefficient between the steel and rubber ball should be the major source of the difference in the energy dissipation between the two systems.     

By incorporating all the above factors, Eq. (\ref{eq:BD2}) can be expressed as follow: 
\begin{eqnarray}
\mathcal{E}=\frac{\partial}{\partial t}T^\ast=-\mathcal{D}-\frac{T^\ast}{\tau}+\frac{T_{b}}{\tau}.  \label{eq:E}
\end{eqnarray}
In a steady state, we should have the relation $\mathcal{E}=0$. 
This relation is just a consequence of energy conservation, simply implying that the energy input to a 2D granular system is partially dissipated by inelastic interparticle collisions and 
by viscous damping due to the particle-wall interactions. 
For an elastic system where $\alpha=1$ and thus $\mathcal{D}=0$, this relation together with $\mathcal{E}=0$ reduces to the relation $T_b=T^\ast$.  
Strictly speaking, we should also consider the energy flux arising from the spatial inhomogeneity of $\phi$ and $T^\ast$  (see below), 
but we tentatively ignore it here since it is not relevant to the description od a steady state.  

On the basis of the above physical picture, we argue that the strong discontinuous nature of the transition of an inelastic system found here 
is a consequence of the inhomogenization of a system under the above-mentioned constraint $\mathcal{E}(\phi,T^\ast)=0$.  
Here we assume that $\mathcal{E}$ is a function of the two state variables, $\phi$ and $T^\ast$.  
First the condition of equal pressure $P(\phi_L,T^\ast_L)=P(\phi_S,T^\ast_S)=P_0$ immediately tells us that $T^\ast_L>T^\ast_S$ since $\phi_L<\phi_S$, 
which is seen in Fig. \ref{fig:coexistence}(d).  
We note that this condition may be robust since it is of mechanical origin, as described before.  
Then the two conditions $\mathcal{E}(\phi_L,T^\ast_L)=\mathcal{E}(\phi_S,T^\ast_S)=0$, together with Eq. (\ref{eq:E}), tell us that the denser solid phase dissipates more energy.  
Another condition is the chemical equilibrium condition, which is also a subtle issue. 
Chemical-balance conditions of two coexisting phases with different temperatures should be obtained as a steady state solution of the relevant kinetic equations 
also considering bond orientational and translational ordering. 
It may be a promising way to introduce a non-equilibrium free energy \cite{argentina2002van}, $\mathcal{F}(\phi,T^\ast,P)$, 
and consider the balance of $\partial \mathcal{F}/\partial \phi$ between the two phases, 
which are expected to be the necessary conditions to maintain a steady state. 
Our discussion is purely phenomenological, and furthermore we also need to consider $\psi_6$. 
Thus, a more rigorous approach is highly desirable in the future. 
  
The lever rule $A_S/A_L=(\phi-\phi_L)/(\phi_S-\phi)$ is then obtained from the condition (1) $N=N_L+N_S$, i.e., $\phi_L A_L+\phi_S A_S=\phi A$, where $\phi_i$, 
$N_i$, and $A_i$, are the particle area fraction, the number of particles, and the total area of phase $i$, and (2) $A=A_L+A_S$. 
The similar lever rule was also observed for the monolayer-bilayer transition 
(see, e.g., \cite{urbach2004,clerc2008liquid}). 
The crucial difference from a thermal system arises from the fact that in our system the granular temperature 
(i.e., the kinetic energy), $T^\ast$, is different between the two phases. The above condition $\mathcal{E}=0$  
is required for maintaining the dynamical steady state dissipating energy and it may be this condition 
that plays the most crucial role in the phase selection in a non-equilibrium state.

\subsection{Dissipation-induced wetting}

The side wall-particle collisions dissipate energy with a rate different from bulk, which affects the wettability of a phase to a solid side wall, as shown in Fig. \ref{fig:boundary}. 
From Eq. (\ref{eq:E}), we can infer that a wall harder than particles tends to wet a liquid phase of high $T^\ast$ rather than a solid phase of low $T^\ast$ 
whereas a wall softer than particles tends to wet a solid phase rather than a liquid phase. 
This is simply because particles near a hard/soft side wall tend to have higher/lower $T^\ast$. 
This spatial inhomogeneity of $T^\ast$ near a side wall should be coupled to the local area fraction $\phi$ to satisfy the following steady state condition for the balance of the heat flux $\mathbf{q}$  \cite{brey1998hydrodynamics}: 
\begin{equation}
\mathbf{q}=-\kappa \nabla T^\ast-\lambda \nabla \phi=0, \label{eq:q}
\end{equation}
where $\kappa$ and $\lambda$ are transport coefficients. 
This condition qualitatively explains why the solid phase with a higher $\phi$ tends to wet to the softer wall, which locally lowers $T^\ast$.

\subsection{Dissipation-induced demixing}

Demixing shown in Fig. \ref{fig:mixture} can be explained by the fact that the energy loss upon collision with the top and bottom plates is 
smaller for steel balls than for rubber balls. The resulting higher kinetic energy of steel balls is a reason why steel balls tends to be located in a liquid phase of hight $T^\ast$. 
This is again consistent with the physical principle discussed above. 
A higher kinetic energy of a steel ball should lead to 
a larger specific area per particle for it. Such a tendency can be seen in Fig. \ref{fig:mixture}(a) and (b).  
This can also be explained by the condition of the balance of the heat flux $\mathbf{q}$ (see Eq. (\ref{eq:q})), 
which tells us the negative correlation between the local effective temperature $T^\ast$ and the local area fraction $\phi$.   

\section{Conclusion and Outlook}
To summarize, we find that energy dissipation plays a crucial role in self-organization 
of driven monolayer granular matter, which allows the coexistence of states with different effective temperatures,  
contrary to the phase coexistence in a thermal system. 
The two-phase coexistence obeys the lever rule as in a thermodynamic first-order transition;  
however, the underlying selection rule is fundamentally different in the sense 
that the energy dissipation rate, which is an intrinsically non-equilibrium quantity, is the key factor of the phase selection. 
Since our discussion is phenomenological, however, it is desirable to theoretically describe the coexistence conditions in a non-equilibrium steady state in a more rigorous manner. 
There are also many fundamental open questions such as what determines the interface profile  
and what is the nature of fluctuations of the interface. 
We also show that it is possible to separate particle species by solidification using the inelasticity contrast, 
which is similar to \emph{purification} of materials including impurities by crystallization in thermodynamic systems. 
Our findings may shed light on a general principle governing the state selection of granular matter far from equilibrium, 
which should also be important for our understanding of industrial processing of granular materials by vibration and flow.   

Finally we expect that a similar principle may hold for the state selection in active matter, which is another important class of out-of-equilibrium systems. 
For active systems, there have recently been many studies on a liquid-solid transition \cite{bialke2012crystallization,fily2012,palacci2013living,goto2015,Li2015}, glass transition \cite{berthier2013non,ni2013pushing,berthier2014nonequilibrium}, demixing  of self-propelled particles \cite{theurkauff2012dynamic,schwarz2012phase,buttinoni2013dynamical,zottl2014hydrodynamics} and rotors \cite{glotzer2014,goto2015,Yeo2015}. 
These studies have elucidated unique characters of the phenomena distinct from their thermodynamic counterparts. 
It has been clarified that the coupling between local energy input (or motility) and density \cite{sokolov2007concentration} plays a crucial role in the state selection (see, e.g., \cite{tailleur2008statistical,farrell2012pattern}).  
Inclusion of inelastic interactions, nonlocal viscous dissipation, and local friction may also alter the nature of 
the state selection in such a system significantly \cite{zottl2014hydrodynamics,goto2015,Li2015}. 
In active matter, and particularly in living systems, interactions between active objects are often dissipative or non-conservative. 
Thus, it is quite interesting and important to study effects of dissipative interactions such as inelasticity and local (or nonlocal) friction 
on the state selection of active matter in a systematic manner. We hope that our study can aid the understanding of fundamental roles of dissipative interactions  
in self-organization of out-of-equilibrium systems.    

\section*{Acknowledgments}
We thank John Russo and Taiki Yanagishima for critical reading of our manuscript. 
This study was partly
supported by Grants-in-Aid for Scientific Research (S) and Specially Promoted Research
from the Japan Society for the Promotion of Science (JSPS).

\section*{Appendix A: Comparison of our study with previous works in quasi-2D systems} 

\subsection*{A-1: Bilayer-forming solidification in quasi-2D driven granular systems}
Here we compare the results of our work with those of previous works done in a quasi-2D situation, 
for which bilayer formation is allowed \cite{urbach2004,urbach2005dynamics,urbach2008effect,urbach2009effects,luu2013capillarylike,clerc2008liquid}.   
All the previous works, which reported liquid-solid coexistence, were made for a cell height $h$ of about 1.7-2.0 times of the particle diameter $d$. Accordingly, these systems can form bilayers, and the liquid-solid transition in these works always accompanies monolayer (liquid)-bilayer (solid) transition. Thus, the situation is essentially different from ours, where the cell thickness is thin enough to avoid bilayer formation. Interestingly, this extra degree of freedom, bilayer formation, makes the physics essentially different from our case in which granular particles always form a monolayer. 

The crucial difference can be seen most clearly in the dependence of the phase behavior on the strength of inelasticity. 
Urbach and his coworkers reported that inelasticity significantly expands the low-density liquid region for a system in which 
bilayer formation is allowed \cite{urbach2008effect,urbach2009effects}. 
Furthermore, computer simulations showed
that the ordered phase is not present at any vibration amplitude when the inelasticity is large. 
Similar behaviors were also reported by Clerc et al. \cite{clerc2008liquid}. 
These results clearly indicate that {\it ordering is suppressed by inelasticity}. 
We stress that this is completely the opposite to our case: In our case inelasticity changes the nature of the transition from continuous to discontinuous  
and the upper bound density of a disordered liquid state is lower for the inelastic system than for the elastic one, i.e., {\it inelasticity helps the ordering}. 
Another important difference comes from the fact that bilayer formation leads to a change in the number density of the bottom layer 
because of the conservation of the total number of particles. This extra degree of freedom also causes 
a difference in the physics between the two types of systems.   

Lobkovsky et al. \cite{urbach2009effects} 
showed by numerical simulations that the monolayer-bilayer transition becomes rather continuous for random forcing. 
This is an interesting observation in the sense that the way of driving affects the nature of the transition. 
However, they also showed that inelasticity suppresses the onset of the ordered phase with random forcing, as is 
observed in the vibrating system. We stress that this tendency is again the opposite to our case. 

Furthermore, we note that the monolayer-bilayer formation can take place as a function of $\Gamma$ \cite{urbach2004,castillo2012fluctuations}, 
whereas our transition \emph{cannot} be induced by changing $\Gamma$, as we can see from the phase diagrams in Fig. \ref{fig:PD}. 
A change from a discontinuous to a continuous nature of the $\Gamma$-induced 
transition was observed with an increase in the normalized cell thickness $h/d$ from 1.83 to 1.94 \cite{castillo2012fluctuations}. 
Here the key control parameter is the cell thickness, which affects the symmetry of the bilayer solid phase (square or hexatic symmetry) 
\cite{urbach2004,castillo2012fluctuations}, and not the degree of inelasticity. 
This fact and the above-mentioned crucial difference in the $\Gamma$-dependence of the transition also
tell us that the physics is essentially different between the two cases. 

The phase ordering behavior in granular matter was proposed to be generally expressed by the Ginzburg-Landau-type free energy (or potential) and  
relevant equations of motion \cite{argentina2002van}. 
On the basis of this picture, a theoretical explanation for bilayer formation was proposed by Clerc et al. \cite{clerc2008liquid}. 
The thermodynamic force on the density field $u$ mainly comes from the effective (non-equilibrium) pressure gradient calculated from this potential. The authors also included the friction term for the $u$ field and random force noises. The kinetic equation governing the phenomena was shown to be non-diffusive and include not only $\partial u/\partial t$ (i.e, $\partial \phi/\partial t$) 
but also $\partial^2 u/\partial t^2$ (i.e.,  $\partial^2 \phi/\partial t^2$) \cite{clerc2008liquid}. 
The sum of these forces leads to acceleration of the field $u$. The reaction-diffusion-type equation of motion with these forces captures the travelling wave features. 
This is markedly different from our monolayer case, where we have never observed such travelling waves.

We confirmed by comparing results of experiments and those of Brownian dynamics simulations 
that our driven monolayer granular system obeys Langevin (or Brownian) dynamics with viscous damping and random noise \cite{watanabe2011}. 
This indicates that there is no acceleration term, which is proportional to $\partial^2 \phi/\partial t^2$, in our system. 
Thus, the above theory cannot be applied to our monolayer case. 
This presence or absence of the acceleration term crucially affects the dynamics and make our monolayer system distinct from 
a bilayer-forming system.  
Nevertheless, it does not influence a steady state, since there $\partial X/\partial t=0$ for any quantity $X$,
In a steady state, thus, the two-phase coexistence is determined by the Ginzburg-Landau-type potential \cite{argentina2002van}. 
This coexistence is then controlled solely by the parameter $\epsilon$, which is a coefficient of the quadratic term ($u^2$) in the potential. 
This parameter $\epsilon$ was assumed to be proportional to the inverse of the compressibility coefficient, as in the case of a usual thermodynamic gas-liquid coexistence.  This (negative) compressibility tells us how easily bilayers can be formed. In this model, the coexistence is determined by the equal pressure and the equal non-equilibrium chemical 
potential under the mass conservation. 
The parameter $\epsilon$ was shown to be given by the derivative of the momentum flux with respect to the density 
and its negative value, which leads to the coexistence, reflects the fact that the granular temperature is lower for a higher density \cite{argentina2002van}. 
Thus, this non-equilibrium chemical potential allows the granular temperature to be different between the two coexisting phases.

Here we propose an intuitive explanation for the role of inelasticity in the liquid-solid transition accompanying bilayer formation, i.e.,  a relation between the above $\epsilon$ and inelasticity. For a quasi-2D system which can form a bilayer, even for elastic particles there is a liquid-solid coexistence (see, e.g., Fig. 3 in \cite{clerc2008liquid}). 
This is because the bilayer formation should be easier for more elastic particles since a smaller loss of the kinetic energy associated with interparticle collisions 
allows particles in the bottom layer to more efficiently jump to the top layer. 
With an increase in the inelasticity, the discontinuous nature becomes weaker and eventually the transition becomes continuous for particles having a small enough restitution coefficient. This termination of the discontinuous first-order-like transition was called a critical point ($\epsilon=0$) by Clerc et al. \cite{clerc2008liquid}.   
Criticality was also observed experimentally, using $\Gamma$ as a control parameter, when $\Gamma$ approaches a critical value \cite{castillo2012fluctuations}. 
The theory may be valid for a quasi-2D driven granular system where bilayer formation is allowed. 
However, we emphasize again that the above-mentioned dependence on the inelasticity is completely opposite to what we find in our system: 
Our central finding is that an increase in inelasticity changes the nature of the 2D liquid-solid transition from (thermal-like) two-step continuous transitions to a one-step discontinuous transition.  
For bilayer formation the kinetic energy has to overcome both gravity and energy loss due to inelastic collisions. This tells us that more elastic particles can form a bilayer more easily, as described above.  To our knowledge, our work is the first to show a discontinuous first-order-like liquid-solid transition of a monolayer granular system in a high $\Gamma$ regime, 
whose nature is primarily controlled by energy dissipation, as discussed in Sec. IV.

\subsection*{A-2: Pattern formation due to inhomogeneization of the energy input}

Next, we mention another `apparently' similar behavior, but which also has a very different physical origin. 
Olafsen and Urbach \cite{olafsen1998clustering} reported clustering or ordering upon decreasing the vertical vibration amplitude for a quasi-2D system, but which has no upper plate and thus is not confined by two parallel walls unlike our case:  
At large $\Gamma$, particle correlations exhibit only short-range order as in the case of equilibrium 2D hard-sphere gases, but 
lowering $\Gamma$ cools the system, resulting in a dramatic increase in correlations leading to either clustering or an ordered
state. Further cooling forms a collapse: a condensate of motionless balls coexisting with a less dense gas. 
Measured velocity distributions are non-Gaussian, showing nearly exponential tails. 
In our systems, we observed similar phenomena below a critical value of $\Gamma_c$, which was shown in Fig. \ref{fig:PD}(b) for a rubber system as the 
``inhomogeneous excitation'' state (the gray region). For a steel ball system this phenomenon was observed for a much lower value of $\Gamma$, as described in the main text. 
We note that these phenomena have an essentially different physical origin, which is for example discussed in detail 
on the basis of molecular dynamics (MD) simulation \cite{nie2000dynamics} and a Navier-Stokes granular hydrodynamics \cite{khain2011hydrodynamics}. 
Nie et al. showed by MD simulation \cite{nie2000dynamics} that at high $\Gamma$ the particle motion
is isotropic, and the velocity distributions are Gaussian. The deviations from a Gaussian 
distribution at low $\Gamma$ is related to the degree of anisotropy in the motion. 
Below $\Gamma_c$, the vertical velocity distribution becomes bimodal: The cluster particles move with the plate, while the gas particles are non-interacting, as they
collide primarily with the plate.  They proposed that dissipative contact forces are responsible for this phenomenon. 
It was also mentioned by Khain and Aranson \cite{khain2011hydrodynamics} that the phenomenon can be viewed as a consequence of a negative compressibility of granular gas. 
This explanation is somewhat similar to that for bilayer formation discussed above \cite{clerc2008liquid}, but 
the range of $\Gamma$ is very different between the two: the clustering is observed for low $\Gamma$, whereas the ordering accompanying bilayer formation for much higher $\Gamma$.  
In relation to this, it was stated \cite{khain2011hydrodynamics} that the behavior does not significantly depend on the inelasticity of collisions between the particles; one does not
need inelastic particle collisions to reproduce experimental observations. They proposed that the mechanism of phase separation occurring at low $\Gamma$ is related to the non-trivial interplay between the energy injection and the vertical temperature of the particles. 
 In any case, thus, an inhomogeneous granular temperature is primarily a consequence of inhomogeneous energy injection.  It was also shown that the behavior does not significantly depend on the inelasticity of collisions between the particles. This clearly indicates that this phenomenon has a physical origin essentially different from ours.

\subsection*{A-3: Gas-liquid coexistence in driven inelastic particles}
Finally, we mention gas-liquid coexistence observed in driven inelastic particles \cite{roeller2011liquid}. In this case, the difference in the type of particle motion between the two phases was found to play a key role in the coexistence: In the dense liquid phase, the injected energy is quickly dissipated within the bulk by frequent interparticle collisions due to a high density. In the dilute gas phase, on the other hand, the motion of particles is synchronized with the driving, which reduces the relative velocity between particles and thus the rate of interparticle collisions. Thus, two phases with a large difference in the granular temperature coexist. However, this phenomenon is also essentially different from ours, reflecting the difference in the cell thickness (many layers vs. monolayer) and the nature of the transition (gas-liquid vs. liquid-solid transition); for example, there is no such synchronized motion in our system.

\section*{Appendix B: On the spatial uniformity of the energy input}

Recently Brito et al.  \cite{brito2013hydrodynamic}  showed that for a granular monolayer vibrated between the walls, uniform energy input to a system is generally a good assumption. In a quasi-2D geometry, in which only a monolayer can exist,  the system is known to remain homogeneous in the horizontal directions for a wide range of parameters. This is due to the presence of a distributed energy source. They also pointed out that in the absence of friction, this energy source is Galilean invariant and conserves momentum locally. In such a quasi-2D system, the vertical energy scale of grains is fixed by vibration parameters. We stress that the energy injection occurs only through direct collisions of a particle with the walls, and there is no other channel. This means that the energy injection rate is controlled by $f_{pw}$. So, as long as there are no direct geometrical restrictions to the vertical motion of the particles, it is reasonable to assume that the energy input is  rather homogeneous spatially. We note that, according to our observation, there is no overlap of particle images projected onto the horizontal plane (see, e.g., images in Fig. \ref{fig:coexistence}). For steel balls, the input energy is homogeneous up to $\phi=0.80$, which is much higher than the upper bound area fraction of  the liquid phase, $\phi_{s}$ (=0.695), of the rubber ball system. Importantly, as shown in Fig. \ref{fig:PD}, all the phase transition area fractions, $\phi_h$, $\phi_s$, $\phi_L$, and $\phi_S$, are independent of $\Gamma$. This suggests that the particle-wall collision frequency $f_{pw}$ may be almost the same between the two phases, or rather homogeneous spatially, for this range of $\Gamma$. We note that if the particle-wall collision frequency $f_{pw}$ (i.e., a vibrational parameter) strongly depends on $\phi$, the phase boundary compositions should also depend on $\Gamma$. Thus, the energy input rate may be assumed to be  homogeneous.  However, since the discussion above is qualitative, the spatial distribution of the energy input in two-phase coexistence needs to be checked carefully by numerical simulations in the future. 

For a quasi-2D system, in which a bilayer can be formed, the situation is very different. For example, it was clearly shown \cite{urbach2009effects} by mapping the local average rate of energy input that a square phase consisting of bilayer corresponds to a region of dramatically reduced energy input. However, this is natural since the bilayer formation itself inevitably accompanies a change in vibrational parameters; for example, the effective mass of the vibrated object is roughly doubled by bilayer formation. We note that such effects are absent for a monolayer system, as mentioned above.   

%

\end{document}